\documentclass[12pt]{article}
\usepackage{amsmath,amssymb}
\usepackage[english]{babel}
\usepackage[cp1251]{inputenc}
\usepackage{hyperref}

\numberwithin{equation}{section}
\newcommand{\bq}{\begin{eqnarray}}
\newcommand{\eq}{\end{eqnarray}}

\newcommand{\bbq}{\begin{equation*}}
\newcommand{\eeq}{\end{equation*}}

\newcommand{\tp}{M}
\newcommand{\tm}{\widetilde m}
\newcommand{\wmu}{\widetilde{\mu}_{\rm x}}

\newcommand{\ra}{\rightarrow}

\newcommand{\ov}{\overline}

\newcommand{\la}{\Lambda_Q}
\newcommand{\lm}{\Lambda_2}
\newcommand{\lt}{\tilde{\Lambda}}
\newcommand{\mx}{\mu_{\rm x}}
\newcommand{\tq}{\textsf{q}}
\newcommand{\otq}{\ov{\textsf{q}}}
\newcommand{\mph}{\mu_{\Phi}}
\newcommand{\lym}{\Lambda_{SYM}}
\newcommand{\mos}{\mu_{o}^{\rm str}}
\newcommand{\w}{{\cal W}_{\rm matter}}
\newcommand{\bb}{2N_c-N_F}

\newcommand{\bo}{{\rm b_o}}
\newcommand{\bd}{{\rm\ov b}_{\rm o}}
\newcommand{\nd}{{\ov N}_c}
\newcommand{\mo}{\mu_{\Phi,\rm 0}}
\newcommand{\qq}{{\langle\ov Q}Q\rangle}

\newcommand{\no}{{\rm n}_1}
\newcommand{\nt}{{\rm n}_2}

\newcommand{\mtl}{\langle m_Q^{\rm tot}\rangle_L}

\newcommand{\Qo}{({\ov Q}Q)_1}
\newcommand{\Qt}{({\ov Q}Q)_2}

\newcommand{\qo}{({\ov q}q)_1}
\newcommand{\qt}{{(\ov q}q)_2}

\newcommand{\qma}{\langle m_{Q,1}^{\rm tot}\rangle}
\newcommand{\qmb}{\langle m_{Q,2}^{\rm tot}\rangle}

\newcommand{\omp}{\mu^{\rm pole}_{q,1}}
\newcommand{\tmp}{\mu^{\rm pole}_{q,2}}

\newcommand{\mgo}{\mu^{\rm pole}_{gl,1}}
\newcommand{\mgt}{\mu^{\rm pole}_{gl,2}}

\newcommand{\wh}{\widehat}

\begin{document}

\begin{center}{\bf \large Mass spectra in $\mathbf{{\cal N}=1\,\, SQCD}$ with additional colorless fields.\\
 Strong coupling regimes.} \end{center}
\vspace{1cm}
\begin{center}{\bf Victor L. Chernyak $^{a,\, b}$} \end{center}
\begin{center}(e-mail: v.l.chernyak@inp.nsk.su) \end{center}
\begin{center} a)\,\, Novosibirsk State University, 630090 Novosibirsk, Russia \end{center}
\begin{center} b)\,\,Budker Institute of Nuclear Physics SB RAS, Novosibirsk, Russia \end{center}
\vspace{1cm}

\begin{center}{\bf Abstract} \end{center}
\vspace{1cm}

This article continues our previous studies of ${\cal N}=1$ SQCD-like theories. We also consider here the $SU(N_c)$ SQCD-like (direct) theory (and its Seiberg's dual with $SU(\nd=N_F-N_c)$ dual colors), with  $N_F$ flavors of light equal mass $m_Q$ quarks ${\ov Q}_j, Q^i,\,\, m_Q\ll\la$. Besides, there are $N_F^2$ additional colorless but flavored fields $\Phi^j_i$ with the large mass parameter $\mph\gg\la$. But now considered in details is the region $N_c<N_F<3N_c/2$
where the UV free direct $SU(N_c)$ theory is strongly coupled at scales $\mu<\la$.

The mass spectra of this direct theory in various vacua and at different values of $\mph$ are calculated within the dynamical scenario introduced by the author in \cite{ch3}. This scenario assumes that quarks in such ${\cal N}=1$ SQCD-like theories without elementary colored adjoint scalars can be in two {\it standard} phases only. These are either the HQ (heavy quark) phase where they are confined or the Higgs phase.

Similarly to our previous studies of this theory within the conformal window $3N_c/2<N_F<3N_c$, it is shown here that due to the strong powerlike RG evolution, the seemingly heavy and dynamically irrelevant at scales $\mu<\la$ fields $\Phi^j_i$ can become light and relevant at lower energies, and there appear then two additional generation of light $\Phi$-particles with masses $\mu_{2,3}^{\rm pole}(\Phi)\ll\la$.

The calculated mass spectra of this strongly coupled at $\mu<\la$ direct $SU(N_c)$ theory were compared to those of its weakly coupled at $\mu<\la$ Seiberg's dual $SU(N_F-N_c)$ variant and appeared to be parametrically different.

All results obtained within the used dynamical scenario from \cite{ch3} look self-consistent. In other words, no internal inconsistencies were encountered in all cases considered. It is worth to remind that the dynamical scenario from \cite{ch3} used in this article satisfies all those tests which were used as checks of the Seiberg hypothesis about the equivalence of the direct and dual theories. This parametrical difference of mass spectra of the direct and dual theories shows, in particular, that all these tests, although necessary, may well be insufficient (see e.g. Appendix A in \cite{session}).
\newpage

\tableofcontents

\newpage

\section{Introduction}

\hspace*{1cm} We continue in this article our previous study in \cite{ch5,ch6,epj} (see also Conclusions in \cite{session}) of ${\cal N}=1$ SQCD-like theories with $SU(N_c)$ colors, $N_F$ flavors of light quarks and with additional $N_F^2$ colorless but flavored fields $\Phi^j_i$. But now considered in details is the region $N_c<N_F<3N_c/2$. It is implied that the reader is familiar with previous author's papers on related subjects. For this reason, we refer frequently below in this text to these previous papers because it is impossible to reproduce once more here all corresponding reasonings and results from these previous author's papers.
\footnote{\,
See in particular more detailed Introduction in \cite{ch6} and Conclusions in \cite{session} on the present status of Seiberg's hypothesis about duality in
${\cal N}=1$ SQCD. Besides, for a reader convenience, we reproduce below from \cite{ch6} some results only for the dual theory.
}

Recall that the Lagrangian of this (direct) $\Phi$-theory at the scale $\mu=\la$ has the form
\footnote{\,
The gluon exponents are always implied in the Kahler terms. Besides, here and everywhere below in the text we neglect for simplicity all RG-evolution effects if they are logarithmic only.
}
\bbq
K={\rm Tr}\, (\Phi^\dagger \Phi )+{\rm Tr}\Bigl (\,Q^\dagger Q+Q\ra{\ov Q}\,\Bigr )\,,\quad {\cal W}=-\frac{2\pi}{\alpha(\mu=\la)}S+{\cal W}_{\rm matter}\,,
\eeq
\bq
{\cal W}_{\rm matter}={\cal W}_{\Phi}+{\cal W}_Q\,,\quad{\cal W}_{\Phi}=\frac{\mph}{2}\Biggl [{\rm Tr}\,(\Phi^2)-\frac{1}{N_F-N_c}\Bigl ({\rm Tr}\,\Phi\Bigr )^2\Biggr ],\quad {\cal W}_Q={\rm Tr}\,\Bigl ({\ov Q}(m_Q-\Phi) Q \Bigr )\,.\label{(1.1)}
\eq

Here\,: $\mph\gg\la$ and $m_Q\ll\la$  are the mass parameters, the traces in \eqref{(1.1)} are over color and/or flavor indices, $S=\sum_{A,\beta} W^{A}_{\beta}W^{A,\,\beta}/32\pi^2$, where $W^A_{\beta}$ is the gauge field strength, $A=1...N_c^2-1,\, \beta=1,2$,\, $a(\mu)=N_c \alpha(\mu)/2\pi=N_c g^2(\mu)/8\pi^2$ is the running gauge coupling with its scale factor $\la$, $\,Q^i_a, {\ov Q}_j^{\,a},\,\,a=1...N_c,\,\,
i,j=1...N_F$ are the quark fields. {\it This normalization of fields is used everywhere below in the main text}. Besides, the perturbative NSVZ $\beta_{NSVZ}$-functions for (effectively) massless SUSY theories \cite{NSVZ1,NSVZ2} are used in this paper.

There is a large number of vacua in this theory and for a reader convenience we reproduce in Appendix the values of the quark and gluino condensates, $\langle{\ov Q}_j Q^i\rangle\equiv\sum_{a=1}^{N_c}\langle{\ov Q}^{\,a}_j Q_a^i\rangle$ and $\langle S\rangle$, in various vacua at $N_c<N_F<2N_c$.

All dynamical properties of theory \eqref{(1.1)}\,: the RG evolution, phase states, mass spectra etc.,  depends essentially on the value of $N_F/N_c$. For instance, it enters at $\mu<\la$\,: a) the weakly coupled IR free logarithmic regime with the gauge coupling $a(\mu\ll\la)\sim 1/\log(\la/\mu)\ll 1$ at $N_F>3N_c$;\, b) the strongly coupled conformal regime with $a(\mu\ll\la)=a_{*}={\rm co nst}=O(1)$ (in general) at $3N_c/2<N_F<3N_c$;\, c) the (very) strongly coupled regime with $a(\mu\ll\la)\sim (\la/\mu)^{\nu_Q\,>\,0}\gg 1,\, \nu=(3N_c-2 N_F)/(N_F-N_c)$ at $N_c<N_F<3N_c/2$. Besides, the mass spectra at given $N_F/N_c$ depend essentially on the considered vacuum.

In parallel with the direct $\Phi$-theory \eqref{(1.1)}, we study also proposed by Seiberg's \cite{S1,S2,IS} its dual variant, the $d\Phi$-theory with $SU(\nd=N_F-N_c)$ dual colors, $N_F$ flavors of dual quarks $q_i^b, {\ov q}^{\,j}_b,\,\,b=1...\nd$, and with $N_F^2$ additional colorless but flavored elementary fields $M^i_j \ra ({\ov Q}_j Q^i)$. Its Lagrangian at $\mu=\la$ looks as, see \eqref{(1.1)} for ${\cal W}_{\Phi}$,
\bbq
{\ov K}={\rm Tr}\,(\Phi^\dagger\Phi)+ {\rm Tr}\,\Bigl (\frac{M^{\dagger}M}{\la^2}\Bigr )+{\rm Tr}\,\Bigl ( q^\dagger q + (q\ra{\ov q})\Bigr )\,,\quad
\ov{\cal W}=\, -\,\frac{2\pi}{{\ov\alpha}(\mu=\la)}\,{\ov S}+{\ov{\cal W}}_{\rm matter}\,,
\eeq
\bq
{\ov{\cal W}}_{\rm matter}={\cal W}_{\Phi}+{\cal W}_{\Phi M}+{\cal W}_q\,,\quad
{\cal W}_{\Phi M}={\rm Tr}\,(m^{\rm tot}_Q M)= {\rm Tr}\,\Bigl ( (m_Q-\Phi)M\Bigr )\,,\quad {\cal W}_q= -\,\rm {Tr}\, \Bigl ({\ov q}\,\frac{M}{\la}\, q \Bigr )\,. \label{(1.2)}
\eq
Here\,:\, the number of dual colors is ${\ov N}_c=(N_F-N_c)$,\, $M^i_j$ are $N_F^2$ Seiberg's elementary mion fields, $M^i_j\ra ({\ov Q}_j Q^i)$,\,\, ${\ov a}(\mu)=\nd{\ov \alpha}(\mu)/2\pi=\nd{\ov g}^2(\mu)/8\pi^2$ is the dual running gauge coupling (with its scale parameter $\Lambda_q$),\,\,${\ov S}= \sum_{B,\beta}{\rm\ov W}^{\,B}_{\beta}\,{\rm \ov W}^{\,B,\,\beta}/32\pi^2$,\,\, ${\rm \ov W}^{\,B}_{\beta}$ is the dual gluon field strength, $B=1...\nd^2-1$. The gluino condensates of the direct and dual theories are matched, $\langle{-\,\ov S}\rangle=\langle S\rangle=\lym^3$, as well as $\langle M^i_j\rangle\equiv\langle M^i_j(\mu=\la)\rangle=\langle{\ov Q}_j Q^i (\mu=\la)\rangle\equiv\langle{\ov Q}_j Q^i\rangle$, and the scale parameter $\Lambda_q$ of the dual gauge coupling is taken as $|\Lambda_q|=\la$. At $N_c<N_F<3N_c/2$ this dual theory is IR free and logarithmically weakly coupled at $\mu<\la,\,\, {\ov a}(\mu\ll\la)\sim 1/{\log(\la/\mu)\ll 1}$.

We studied these theories \eqref{(1.1)},\eqref{(1.2)} in previous articles \cite{ch5,ch6,epj} at values of $N_F$ in the range $3N_c/2<N_F<3N_c$, i.e. within the conformal window. The purpose of this article is to consider in details the range $N_c<N_F<3N_c/2$.\\ 

According to Seiberg's view of the standard direct (i.e. without fields $\Phi^i_j$)\, ${\cal N}=1$ SQCD at
$N_c+1<N_F<3N_c/2$, with the scale factor $\la$ of $SU(N_c)$ gauge coupling and direct quarks with $m_Q=0$ (or with $m_Q\ll\Lambda$), the regime of the direct theory at $\mu<\la$ is in this case: {\bf `confinement without chiral symmetry breaking'} (as far as small $m_Q\neq 0$ can be neglected). And {\bf the dual theory is considered as the lower energy form of the direct theory}. This means that all direct quarks remained massless (or light), but hadrons made from these massless (or light) quarks and direct gluons {\bf acquired large masses $\sim\la$ due to mysterious confinement with the string tension} $\sigma^{1/2}\sim\la$, and decoupled at $\mu<\la$. Instead of them, there mysteriously appeared massless (or light) composite solitons. These last are particles of the dual theory.\\

This picture was questioned in \cite{ch1} (see section 7 therein). It was argued that, with the unbroken chiral flavor symmetry $SU(N_F)_L\times SU(N_F)_R$ and unbroken R-charge, it is impossible to write at $\mu\sim\la$ the nonsingular superpotential of the effective Lagrangian of massive flavored hadrons with masses $\sim\la$ made from direct massless (or light) quarks.~
\footnote{\,
This is similar to our ordinary QCD with the scale factor $\Lambda$ of the $SU(3)$ gauge coupling. Then, with confinement and with three massless (or light) quarks, but without breaking of chiral symmetry $SU(3)_L\times SU(3)_R$, it is impossible e.g. to have in the effective hadron Lagrangian at $\mu\sim\Lambda$ the massive nucleons with the mass $\sim\Lambda$, as the term $\sim\Lambda {\ov N} N$ in the potential is incompatible with the unbroken chiral symmetry. And the situation in ${\cal N}=1$ SQCD is even more restrictive because the superpotential is holomorphic and due to additional R-charge conservation.
}

We also recall here the following. There is no confinement in Yukawa-like theories without gauge interactions. The confinement originates {\bf only} from the (part of) YM or ${\cal N}=1$ SYM in ${\cal N}=1$ SQCD-like theories unbroken by (possibly) higgsed quarks. And because ${\cal N}=1$ SYM has only one dimensional parameter $\langle\lym\rangle=\langle S\rangle^{1/3}$, the string tension is $\sigma^{1/2}\sim\langle\lym\rangle$. But in the standard ${\cal N}=1$ SQCD the value of $\lym$ is well known: $\lym=(\Lambda_Q^\bo\det m_Q)^{1/3N_c}\ll\Lambda_Q$. Therefore, such SYM cannot produce confinement with the string tension $\sim\Lambda_Q$ (and there is no confinement at all at $m_Q\ra 0$).~
\footnote{\,
And the same for the direct SQCD-like $\Phi$-theory considered here:\, $\langle\lym\rangle=(\la^{\bo}\det\langle m^{\rm tot}_Q\rangle)^{1/3N_c}\ll\la$. Therefore, such SYM cannot produce confinement with $\sigma^{1/2}\sim\la$, only with $\sigma^{1/2}\sim\langle\lym\rangle$.
}

For these reasons, as was argued in detail in section 7 of \cite{ch1} and in Conclusions of \cite{session}, the UV free direct theory \eqref{(1.1)} with $N_c+1<N_F<3N_c/2$ flavors of light quarks enters smoothly at $\mu<\la$ into the effectively massless perturbative (very) strongly coupled regime with $a(\mu\ll\la)\sim (\la/\mu)^{\nu_Q\,>\,0}\gg 1$ (and NSVZ $\beta$-function allows this). This is qualitatively similar to situation in the conformal window at $3N_c/2<N_F<3N_c$, where the direct theory enters smoothly at $\mu<\la$ into the conformal regime with the coupling $a(\mu\ll\la)={\rm const}\sim 1$, and with all its light quarks and gluons remaining effectively massless.

For a description of the RG evolution and calculations of mass spectra in various vacua in this strong coupling regime we use the dynamical scenario introduced in \cite{ch3}. This scenario assumes that quarks in such ${\cal N}=1$ SQCD-like theories can be in two {\it standard} phases only. These are either the HQ (heavy quark) phase where $\langle{\ov Q}\rangle=\langle Q \rangle=0$ and they are confined, or the Higgs phase where they form nonzero coherent condensate $\langle{\ov Q}\rangle=\langle Q \rangle\neq 0$ breaking the color symmetry. The word {\it standard} implies here also that, unlike e.g. very special ${\cal N}=2$ SQCD with enhanced supersymmetry, in such ${\cal N}=1$ theories without elementary colored adjoint scalars, no {\it additional}
\footnote{\,
i.e. in addition to the massless Nambu-Goldstone particles due to spontaneously broken global flavor symmetry
}
parametrically light solitons (e.g. magnetic monopoles or dyons) are formed at those scales where quarks decouple as heavy or are higgsed.

Within this scenario, we calculate the mass spectra of the strongly coupled direct theory \eqref{(1.1)} in different vacua. It is shown in this paper (as well as in \cite{ch3,ch5,ch6,epj}\,) that the use of this scenario leads to the results for the mass spectra which do not contradict to any proven results and look self-consistent, i.e. no internal inconsistences are incountered in all cases considered.

Similarly to our previous studies of this theory \eqref{(1.1)} within the conformal window  at $3N_c/2<N_F<3N_c$ in \cite{ch5,ch6,epj}, it is shown here that, due to a strong powerlike RG evolution at scales $\mu<\la$ in the direct theory, the seemingly heavy and dynamically irrelevant fields $\Phi^j_i$ with $\mph\gg\la$ can become light and there appear then two additional generations of light $\Phi$-particles with $\mu_{2,3}^{\rm pole}(\Phi)\ll\la$.

In parallel, we calculate the mass spectra in the Seiberg's dual theory \eqref{(1.2)}. But, as explained above, this "dual"\, theory has to be considered {\bf not} as the low energy form of the direct theory at scales $\mu<\la$, but simply as {\bf a definite independent theory}. This IR free theory with $N_c<N_F<3N_c/2$ is logarithmically weakly coupled at $\mu<\la$ and so needs no additional dynamical assumptions for calculations of mass spectra. The mass spectra of both direct and dual theories are then calculated and compared.

As described below in the text, the comparison of mass spectra of direct and dual theories \eqref{(1.1)} and \eqref{(1.2)} shows, similarly to the standard direct and dual ${\cal N}=1$ SQCD theories (i.e. those in \eqref{(1.1)} and \eqref{(1.2)} but without fields $\Phi^j_i$, see \cite{ch3}), that these two mass spectra are {\bf parametrically different}. It is worth to recall that the dynamical scenario from \cite{ch3} used in this article satisfies all those tests which were used as checks of the Seiberg hypothesis about the equivalence of the direct and dual theories. This shows, in particular, that all these tests, although necessary, may well be insufficient.

The paper is organized as follows. Because the global flavor symmetry $U(N_F)$ is unbroken or broken spontaneously as $U(N_F)\ra U(\no)\times U(\nt)$ in different vacua of \eqref{(1.1)},\eqref{(1.2)}, we consider these cases separately. Besides, because the parametric behavior of quark condensates and the whole dynamics are quite different in two regions $\mph\gtrless\mo=\la(\la/m_Q)^{(2N_c-N_F)/N_c}\gg\la$, we also consider these regions separately. In the region $\la\ll\mph\ll\mo$, the vacua with unbroken $U(N_F)$ are considered in sections 2-3, while those with broken $U(N_F)$ in sections 4-6 (in both theories \eqref{(1.1)} and \eqref{(1.2)} separately). In the region $\mo\ll\mph\ll\la^2/m_Q$, the vacua with unbroken $U(N_F)$ are considered in section 7, while those with broken $U(N_F)$ in sections 8-9.
\vspace*{1mm}

\addcontentsline{toc}{section}
 {\hspace*{3cm} The region $\la\ll\mph\ll\mo=\la (\la/m_Q)^{(2N_c-N_F)/N_c}$}

\begin{center}{\Large\bf The region $\la\ll\mph\ll\mo=\la (\la/m_Q)^{(2N_c-N_F)/N_c}$} \end{center}

\vspace*{-7mm}

\section{Unbroken flavor symmetry, \,  L-vacua}

\subsection{\quad Direct theory}

The quark condensates (here and everywhere below always at the scale $\mu=\la$) in these L (large)-vacua with the multiplicity $(2N_c-N_F)$ look in this theory \eqref{(1.1)} as (see Appendix, $\,\langle S\rangle$ is the gluino condensate summed over all its colors, $\bo=3N_c-N_F$)
\bq
\qq_L\equiv\langle{\ov Q} Q(\mu=\la)\rangle_{L}\sim\la^2\Bigl (\frac{\la}{\mph}\Bigr )^{\frac{N_F-N_c}{2N_c-N_F}}\ll\la^2,\,\, \langle S\rangle_L=\Bigl (\frac{ \langle\det{\ov Q}Q\rangle_L}{\la^{\rm \bo}}\Bigr )^{\frac{1}{N_F-N_c}}\sim\la^3\Bigl (\frac{\la}{\mph}\Bigr )_{,}^{\frac{N_F}{2N_c-N_F}}\,\,\label{(2.1)}
\eq
while from the Konishi anomaly \cite{Konishi}
\bq
\mtl\equiv\langle m_Q-\Phi\rangle_L=\frac{\langle S\rangle_L}{\qq_L}\sim\la\Bigl (\frac{\la}{\mph}\Bigr )^{\frac{N_c}{2N_c-N_F}}\ll\la\,.\label{(2.2)}
\eq

As was argued in detail in section 7 of \cite{ch1} and in Conclusions of \cite{session}, this UV free direct $\Phi$-theory with the gauge coupling $ a(\mu\gg\la)\equiv N_c g^2(\mu\gg\la)/8\pi^2\ll 1$ enters smoothly the strong coupling regime at the scale $\mu<\la$, with the gauge coupling $ a(\mu\ll\la)\gg 1$. The values of anomalous dimensions $\gamma_Q$ and $\gamma_q$ of the direct and dual quarks are related at $N_c<N_F<3N_c/2$ and $\mu\ll\la$ as \cite{ch1}
\bbq
\nd (1+\gamma_Q)=N_c (1+\gamma_q)\,,\quad\quad \gamma_q\,\,\,{\xrightarrow {\mu\ll\la}}\,\,\, 0\,,\quad\quad \gamma_Q\,\,\ra\,\, \frac{2N_c-N_F}{N_F-N_c}\,.
\eeq

The potentially important masses look then as follows, see e.g. Appendix in \cite{ch6} and section 5 in \cite{epj} for the values of quark and fion anomalous dimensions, $\gamma_Q$ and $\gamma_{\Phi}= -2\gamma_Q$, at $a(\mu)\gg 1$\,:\\
a) the quark pole mass
\bq
m^{\rm pole}_{Q,L}=\frac{\mtl}{z_Q(\la,m^{\rm pole}_{Q,L})}\sim\la\Bigl (\frac{\la}{\mph}\Bigr )^{\frac{N_F-N_c}{2N_c-N_F}}\,,\,\,\, z_Q(\la,m^{\rm pole}_{Q,L})=\Bigl (\frac{\la}{m^{\rm pole}_{Q,L}}\Bigr )^{\gamma_Q},\,\,\,\gamma_Q=\frac{2N_c-N_F}{N_F-N_c}>\,1\,; \label{(2.3)}
\eq
b) the gluon mass due to possible higgsing of quarks
\bq
\Bigl (\mu^{\rm pole}_{{\rm gl},L}\Bigr )^2\sim a(\mu^{\rm pole}_{{\rm gl},L})\,z_Q(\la,\mu^{\rm pole}_{{\rm gl},L})\qq_L\,,\quad z_Q(\la,\mu^{\rm pole}_{{\rm gl},L})=\Bigl (\frac{\mu^{\rm pole}_{{\rm gl},L}}{\la}\Bigr )^{\gamma_Q}\ll 1,\,\, \label{(2.4)}
\eq
\bbq
\frac{d\,a(\mu\ll\la)}{d\log\mu}=\beta_{NSVZ}(a)=\frac{a^2(\mu)}{a(\mu)-1}\,\frac{\bo-N_F\gamma_Q}{N_c}\quad
\xrightarrow{a(\mu)\gg 1}\quad -\,\nu_Q\,a(\mu),\quad \bo=3N_c-N_F\,,
\eeq
\bbq
\nu_Q=\frac{N_F\gamma_Q-\bo}{N_c}=\frac{3N_c-2N_F}{N_F-N_c}=\gamma_Q-1>0\,,\quad
a(\mu=\mu^{\rm pole}_{{\rm gl},L}\ll\la)\sim\Bigl (\frac{\la}{\mu^{\rm pole}_{{\rm gl},L}}\Bigr )^{\nu_Q}\gg 1,
\eeq
\bbq
a(\mu^{\rm pole}_{{\rm gl},L})\,z_Q(\la,\mu^{\rm pole}_{{\rm gl},L})\sim\frac{\mu^{\rm pole}_{{\rm gl},L}}{\la},\quad \mu^{\rm pole}_{{\rm gl},L}\sim \frac{\qq_L}{\la}\sim\la\Bigl (\frac{\la}{\mph}\Bigr )^{\frac{N_F-N_c}{2N_c-N_F}}\sim m^{\rm pole}_{Q,L}\,.
\eeq
Because the global non-Abelian flavor symmetry $SU(N_F)$ is unbroken in these L-vacua, this means that {\it the quarks in such a case are not higgsed due to the rank restriction} $N_c<N_F$, as otherwise the global flavor symmetry $SU(N_F)$ will be broken spontaneously. Therefore, {\bf the overall phase is in this case HQ} (heavy confined quarks).
\footnote{\,
The same reasonings are used also everywhere below in similar cases. \label{f4}
}

The Lagrangian at scales $\mu$ such that $\mos<\mu<\la$ looks as
\bq
K=z_{\Phi}(\la,\mu)\,{\rm Tr}\, (\,\Phi^\dagger\Phi\,)+z_Q(\la,\mu)\,{\rm Tr}\,\Bigl (Q^\dagger Q+Q\ra \bar Q\Bigr )\,,\quad z_{\Phi}(\la,\mu)=1/z^2_Q(\la,\mu)\,,\label{(2.5)}
\eq
\bbq
z_Q(\la,\mu\ll\la)=\Bigl (\frac{\mu}{\la}\Bigr )^{\gamma_Q}=\Bigl (\frac{\mu}{\la}\Bigr )^{\frac{\bb}{N_F-N_c}}\ll 1,\,\, z_{\Phi}(\la,\mu)=\Bigl (\frac{\mu}{\la}\Bigr )^{\gamma_{\Phi}=-2\gamma_Q}=\Bigl (\frac{\la}{\mu}\Bigr )^{\frac{2(2N_c-N_F)}{N_F-N_c}}\gg 1.
\eeq
\bq
\w={\cal W}_{\Phi}+{\rm Tr}\,(\,{\ov Q}\,m_Q^{\rm tot} Q\,)\,,\quad {\cal W}_{\Phi}=\frac{\mph}{2}\Bigl ({\rm Tr}\,(\Phi^2)-\frac{1}{\nd}({\rm Tr}\,\Phi)^2\Bigr ),\quad m_Q^{\rm tot}=(m_Q-\Phi)\,.\label{(2.6)}
\eq
Therefore, the running perturbative mass of $\Phi$ is $\mu_{\Phi}(\mu\ll\la)=\mph/z_{\Phi}(\la,\mu)\ll\mph$ and, if nothing prevents, the field $\Phi$ becomes dynamically relevant at scales $\mu<\mu_o^{\rm str},\,\mu_{\Phi}(\mu=\mu_o^{\rm str})=\mu_o^{\rm str}$,
\bbq
\mu_o^{\rm str}=\la\Bigl (\frac{\la}{\mph}\Bigr )^{\frac{1}{2\gamma_Q-1}}=\la\Bigl (\frac{\la}{\mph}\Bigr )^{\frac{\nd}{5N_c-3N_F}}\,,\quad \gamma_Q=\frac{2N_c-N_F}{\nd}\,,\quad \nd=N_F-N_c\,,
\eeq
\bq
\frac{\mu_o^{\rm str}}{m^{\rm pole}_{Q,L}}\sim\Bigl (\frac{\mph}{\la}\Bigr )^{\Delta}\gg 1\,,\quad \Delta=\frac{\nd(3N_c-2N_F)}{(2N_c-N_F)(5N_c-3N_F)}>0\,,\label{(2.7)}
\eq
and there is the second generation of all $N_F^2$ fions $\Phi^j_i$ with $\mu_{2,L}^{\rm pole}(\Phi)\sim\mu_o^{\rm str}\gg m_{Q,L}^{\rm pole}$, see section 5 in \cite{ch4}. Besides, even at lower scales $m^{\rm pole}_{Q,L}\ll\mu\ll\mu_o^{\rm str}$ where the fields $\Phi^j_i$ became already effectively massless (in the sense $\mu_{\Phi}(\mu)\ll\mu$ ) and so dynamically relevant, the quark anomalous dimension $\gamma_Q=(\bb)/\nd$ remains the same, as well as $\gamma_{\Phi}=-2\gamma_Q$, see Appendix in \cite{ch6}.

At $\mu<m^{\rm pole}_{Q,L}$ all quarks decouple as heavy ones and the RG evolution of all fields $\Phi$ becomes frozen, but this happens in the region where they are already relevant, i.e. the running mass of fions $\mu_{\Phi}(\mu)$ is $\mu_{\Phi}(\mu=m^{\rm pole}_{Q,L})\ll m^{\rm pole}_{Q,L}$. This means that there is the third generation of all $N_F^2$ fions with $\mu^{\rm pole}_{3,L}(\Phi)\ll m^{\rm pole}_{Q,L}$, see section 5 in \cite{ch4} and this text below.

The lower energy theory at $\mu<m^{\rm pole}_{Q,L}$ is ${\cal N}=1\,\, SU(N_c)$ SYM in the strong coupling regime (plus $N_F^2$ colorless fions $\Phi$). The scale factor $\langle\lym\rangle_L$ of its gauge coupling is determined from the matching, see \eqref{(2.3)},\eqref{(2.4)} and section 7 in \cite{ch1}\,:
\bbq
\frac{d\, a^{\rm str}_{YM}(\mu)}{d\log\mu}=\beta^{YM}_{NSVZ}(a^{\rm str}_{YM})=\frac{3( a^{\rm str}_{YM})^2}{a^{\rm str}_{YM}-1}\,\,\,\,\xrightarrow{a^{\rm str}_{YM}\gg 1}\,\,\, 3\, a^{\rm str}_{YM}(\mu)\,, \quad \,a^{\rm str}_{YM}(\mu\gg\lym)\sim\Bigl ( \frac{\mu}{\lym}\Bigr )^3\gg 1,
\eeq
\bq
a_{+}(\mu=m^{\rm pole}_{Q,L}\ll\la)\sim\Bigl (\frac{\la}{m^{\rm pole}_{Q,L}}\Bigr )^{\nu_Q}=a^{\rm str}_{YM}
(\mu=m^{\rm pole}_{Q,L}\gg\lym)\sim\Bigl (\frac{m^{\rm pole}_{Q,L}}{\langle\lym^{(L)}\rangle}\Bigr )^3\gg 1\,.\label{(2.8)}
\eq
From \eqref{(2.3)},\eqref{(2.4)},\eqref{(2.8)}
\bq
\langle\lym^{(L)}\rangle^{3}\sim\la^3\Bigl (\frac{\la}{\mph}\Bigr )^{\frac{N_F}{2N_c-N_F}}\,,\quad
\frac{\langle\lym^{(L)}\rangle}{m^{\rm pole}_{Q,L}}\sim\Bigl (\frac{\la}{\mph}\Bigr )^{\frac{3N_c-2N_F}{3(2N_c-N_F)}}\ll 1\,,\label{(2.9)}
\eq
as it should be because, see \eqref{(2.1)},
\bbq
\langle\lym^{(L)}\rangle^{3}\equiv\langle S\rangle_L=\Biggl (\frac{\det \langle{\ov Q}Q\rangle_L}{\la^{\rm \bo}}\Biggr )^{1/\nd}\sim \la^3\Bigl (\frac{\la}{\mph}\Bigr )^{\frac{N_F}{2N_c-N_F}}.
\eeq

After lowering the scale down to $\mu<\langle\lym^{(L)}\rangle$ and integrating out all gauge degrees of freedom via the VY (Veneziano-Yankielowicz)-procedure \cite{VY,TVY}, the low energy Lagrangian looks as, see \eqref{(2.2)},\eqref{(2.3)},\eqref{(2.6)}
\bq
K=z_{\Phi}(\la,m^{\rm pole}_{Q,L})\,{\rm Tr}\,(\,\Phi^\dagger\Phi\,)\,,\quad
z_{\Phi}(\la,m^{\rm pole}_{Q,L})=\frac{1}{z^2_Q(\la,m^{\rm pole}_{Q,L})}=\Bigl (\frac{\la}{m^{\rm pole}_{Q,L}}\Bigr )^{\frac{2(2N_c-N_F)}{N_F-N_c}}\gg 1,\label{(2.10)}
\eq
\bbq
{\cal W}={\cal W}_{\Phi}+{\cal W}_{\rm non-pert}\,,\quad {\cal W}_{\rm non-pert}=N_c\Bigl (\lym^{(L)}\Bigr )^3=N_c\Bigl (\la^{\rm \bo}\det m^{\rm tot}_Q\Bigr )^{1/N_c}\,.
\eeq
From \eqref{(2.10)}, the pole masses of $N_F^2$ third generation fions $\Phi^j_i$ are (the contributions to $\mu^{\rm pole}_{3,L}(\Phi)$ from ${\cal W}_{\Phi}$ and ${\cal W}_{\rm non-pert}$ are parametrically the same)
\bq
\mu^{\rm pole}_{3,L}(\Phi)\sim\frac{\mph}{z_{\Phi}(\la,m^{\rm pole}_{Q,L})}\sim\frac{\la^2}{\mph}\,,\quad
\quad\frac{\mu^{\rm pole}_{3,L}(\Phi)}{\langle\lym^{(L)}\rangle}\sim\Biggl (\frac{\la}{\mph}\Biggr )^{\frac{2(3N_c-2N_F)}{3(2N_F-N_c)}}\ll 1\,.\label{(2.11)}
\eq

On the whole for the case considered.\\
1)\,\, All quarks ${\ov Q}_j, Q^i$ are in the HQ (heavy quark) phase and weakly confined (i.e. the tension of the confining string originating from ${\cal N}=1\,\, SU(N_c)$ SYM is much smaller than quark masses, $\sqrt\sigma\sim\langle\lym^{(L)}\rangle\ll m^{\rm pole}_{Q,L}$,  see \eqref{(2.3)},\eqref{(2.9)}.\\
2)\,\, There is a large number of $SU(N_c)$ gluonia with the mass scale $\sim\langle\lym^{(L)}\rangle=\langle S\rangle^{1/3}_{L}$, see \eqref{(2.9)}.\\
3) There are two generations of $N_F^2$ fions $\Phi^j_i$ with masses
\bq
\mu^{\rm pole}_{2,L}(\Phi)\sim\mu_o^{\rm str}\sim\la\Bigl(\frac{\la}{\mph}\Bigr)^{\frac{\nd}{5N_c-3N_F}},\quad \mu^{\rm pole}_{3,L}(\Phi)\sim\frac{\la^2}{\mph}\,.\label{(2.12)}
\eq
The overall mass hierarchies look as
\bq
\mu^{\rm pole}_{3,L}(\Phi)\ll\langle\lym^{(L)}\rangle\ll m^{\rm pole}_{Q,L}\ll\mu^{\rm pole}_{2,L}(\Phi)\ll\la\ll\mu^{\rm pole}_{1}(\Phi)\sim\mph\,.\label{(2.13)}
\eq

\subsection{\quad Dual theory}

The mass spectra in the Seiberg dual IR free and logarithmically weakly coupled at $\mu<\la$ $SU(\nd)$ theory \eqref{(1.2)} were described  for this case in section 7.1 of \cite{ch6}. For the reader convenience and completeness we reproduce here the results. \\
1) All $N_F^2$ fions $\Phi^j_i$ have large masses $\mu_1^{\rm pole}(\Phi)\sim\mph\gg\la$ (with logarithmic accuracy) and are dynamically irrelevant at all lower scales.\\
2) All dual quarks ${\ov q}^j, q_i$ are in {\bf the overall Hq} (heavy quark) phase. There is a large number of hadrons made of weakly interacting non-relativistic and weakly confined dual quarks. The scale of their masses, neglecting logarithmic RG-evolution factors, is $\mu^{\rm pole}_{q,L}\sim \langle M\rangle_{\rm L}/\la=\qq_{\rm L}/\la\sim m^{\rm pole}_{Q,L}$, see \eqref{(2.3)} (the tension of the confining string originating from ${\cal N}=1\,\, SU(\nd)$ SYM is much smaller, $\sqrt \sigma\sim\langle\lym^{(L)}\rangle\ll\mu^{\rm pole}_{q,L}$, see \eqref{(2.9)}\,).  \\
3) A large number of gluonia made of $SU(\nd)$ gluons with their mass scale $\sim\langle\lym^{(L)}\rangle=\langle S\rangle_L^{1/3}\sim \la (\la/\mph)^{N_F/3(2N_c-N_F)}$.\\
4) $N_F^2$ Seiberg's mions $M^i_j$ with masses $\mu^{\rm pole}_{L}(M)\sim\la^2/\mph$.\\

The mass hierarchies look here as $\mu^{\rm pole}_{L}(M)\ll\langle\lym^{(L)}\rangle\ll\mu^{\rm pole}_{q}\ll\la\ll\mu^{\rm pole}_{1}(\Phi)\sim\mph$.\\

Comparing the mass spectra of the direct and dual theories we note the following.\\
1) The weakly confined dual quarks ${\ov q}^j, q_i$ inside dual hadrons are non-relativistic and parametrically weakly coupled (the dual coupling ${\ov a}$ at the scale of the Bohr momentum is logarithmically small). Therefore, the Coulomb mass splittings of the low lying hadrons are also parametrically small, i.e. $\delta M_{H}/M_{H}\sim {\ov a}^{\,2}\ll 1$. There is nothing similar in the direct theory with the strongly coupled quarks ${\ov Q}_j, Q^i$.\\
2) In the range of scales $\mu_o^{\rm str}\ll\mu\ll\la$ the effectively massless flavored particles in the direct theory are only quarks ${\ov Q}_j, Q^i$, while all $N_F^2$ fions $\Phi^j_i$ have large running masses $\mu_{\Phi}(\mu)>\mu$ and are dynamically irrelevant. In the dual theory the effectively massless flavored particles are the dual quarks ${\ov q}^j, q_i$ and $N_F^2$ mions $M^i_j$.  Therefore, the anomalous 't Hooft triangles $SU^3(N_F)_L$ are the same in the direct and dual theories \cite{S2}.\\
3) In the range of scales $\mu^{\rm pole}_{q,L}\sim m^{\rm pole}_{Q,L}\ll\mu\ll\mu_o^{\rm str}$ the effectively massless flavored particles in the dual theory remain the same, while $N_F^2$ fions become effectively massless (i.e. the running mass $\mu_{\Phi}(\mu)$ of all fions is $\mu_{\Phi}(\mu)\ll\mu$\,) and give now additional contributions e.g. to $SU^{\,3}(N_F)_L$ triangles in the direct theory. Therefore, the $SU^{\,3}(N_F)_L$ triangles do not match now in the direct and dual theories.\\
4) At scales $\mu\ll m^{\rm pole}_{Q,L}\sim\mu^{\rm pole}_{q,L}$ all direct and dual quarks decouple. In the range of scales $\mu^{\rm pole}_{3,L}(\Phi)\sim \mu^{\rm pole}_{L}(M)\ll\mu\ll m^{\rm pole}_{Q,L}$ the effectively massless flavored particles in the direct theory are $N_F^2$ third generation fions $\Phi^j_i$, while in the dual theory these are $N_F^2$ mions $M_j^i$. Therefore, the values of $SU^{\,3}(N_F)_L$ triangles differ in sign in the direct and dual theories.

\section{Unbroken flavor symmetry,\,  S-vacua}

\subsection{\quad Direct theory}

The quark condensates in these S (small)-vacua look as, see Appendix,
\bq
\qq_S\simeq -\frac{N_c}{\nd}\, m_Q\mph\,,\quad \langle S\rangle_S=\Bigl (\frac{\det \qq_S}{\la^{\rm \bo}}\Bigr )^{1/\nd}\sim\la^3\Bigl (\frac{m_Q\mph}{\la^2}\Bigr )^{N_F/\nd}\,.\label{(3.1)}
\eq

The direct theory is strongly coupled at $\mu<\la$. Proceeding as in section 2.1 we obtain for this S-vacuum, see \eqref{(2.3)},\eqref{(2.4)},\eqref{(2.8)},\eqref{(3.1)},
\bq
\langle m_Q^{\rm tot}\rangle_S=\frac{\langle S\rangle_S}{\qq_S}\sim \la\Bigl (\frac{\qq_S}{\la^2}\Bigr )^{N_c/\nd}\,,\quad m_{Q,S}^{\rm pole}=\frac{\langle m_Q^{\rm tot}\rangle_S}{z_Q(\la,m_{Q,S}^{\rm pole})}
\sim \frac{\qq_S}{\la}\sim\frac{m_Q\mph}{\la}\ll\la\,,\label{(3.2)}
\eq
\bbq
\Bigl (\mu^{\rm pole}_{{\rm gl},S}\Bigr )^2\sim a(\mu^{\rm pole}_{{\rm gl},S})z_Q(\la,\mu^{\rm pole}_{{\rm gl},S})\,\qq_S\sim\frac{\mu^{\rm pole}_{{\rm gl},S}}{\la}\,\qq_S\quad\ra\quad
\mu^{\rm pole}_{{\rm gl},S}\sim \frac{\qq_S}{\la}\sim\frac{m_Q\mph}{\la}\sim m_{Q,S}^{\rm pole}\,.
\eeq
Therefore, for the same reasons of the rank restrictions and unbroken global flavor symmetry as in L-vacua in section 2.1, {\bf the overall phase is HQ}.
Besides,
\bq
\frac{m_{Q,S}^{\rm pole}}{\mu_o^{\rm str}}\sim\frac{m_Q}{\la}\Bigl (\frac{\mph}{\la}\Bigr )^{\frac{2(2N_c-N_F)}{5N_c-3N_F}}\ll\frac{m_Q}{\la}\Bigl (\frac{\mo}{\la}\Bigr )^{\frac{2(2N_c-N_F)}{5N_c-3N_F}}\sim\Bigl (\frac{m_Q}{\la}\Bigr )^{\frac{\nd(3N_c-2N_F)}{N_c(5N_c-3N_F)}}\ll 1,\,\, \la\ll\mph\ll\mo\,,\label{(3.3)}
\eq
so that the running mass of all fions $\Phi^j_i$ is $\mu_{\Phi}(\mu<\mu_o^{\rm str})<\mu$, all $N_F^2$ fions become dynamically relevant at $\mu<\mu_o^{\rm str}$ and there is the second generation of fions with $\mu_{2,S}^{\rm pole}\sim\mu_o^{\rm str}=\la(\la/\mph)^{\nd/(5N_c-3N_F)}\gg m_{Q,S}^{\rm pole}$.

At $\mu<m_{Q,S}^{\rm pole}$ all quarks decouple as heavy and, proceeding as in \eqref{(2.8)}, we obtain the scale factor of remained ${\cal N}=1\,\, SU(N_c)$ SYM
\bq
a_{+}(\mu=m^{\rm pole}_{Q,S})=\Bigl (\frac{\la}{m^{\rm pole}_{Q,S}}\Bigr )^{\nu_Q=(3N_c-2N_F)/\nd}=a^{\rm str}_{YM}(\mu=m^{\rm pole}_{Q,S})=\Bigl (\frac{\mu=m^{\rm pole}_{Q,S}}{\langle\lym^{(S)}\rangle}\Bigr )^3\gg 1\,.\label{(3.4)}
\eq
From \eqref{(3.2)},\eqref{(3.4)}
\bq
\langle\lym^{(S)}\rangle^{3}\sim\la^3\Bigl (\frac{m_Q\mph}{\la^2}\Bigr )^{N_F/\nd}\,,\quad
\frac{\langle\lym^{(S)}\rangle}{m^{\rm pole}_{Q,S}}\sim\Bigl (\frac{m_Q\mph}{\la^2}\Bigr )^{(3N_c-2N_F)/3\nd}\ll\Bigl (\frac{m_Q}{\la}\Bigr )^{(3N_c-2N_F)/3N_c}\ll 1\,,\label{(3.5)}
\eq
as it should be because, see \eqref{(3.1)},
\bq
\langle\lym^{(S)}\rangle^{3}\equiv\langle S\rangle_S=\Biggl (\frac{\det \langle{\bar Q}Q\rangle_S}{\la^{\rm \bo}}\Biggr )^{1/\nd}\sim\la^3\Bigl (\frac{m_Q\mph}{\la^2}\Bigr )^{N_F/\nd}\,.\label{(3.6)}
\eq

The low energy Lagrangian at $\mu<\langle\lym^{(S)}\rangle$ has the same form as in \eqref{(2.10)}, but now in S - vacua. In this case the main contribution to the masses of third generation fions originates from the nonperturbative term in \eqref{(2.10)} and is
\bq
\mu^{\rm pole}_{3,S}(\Phi)\sim\frac{1}{z_{\Phi}(\la,m_{Q,S}^{\rm pole})}\,\frac{\langle S\rangle_S}{\langle m_{Q,S}^{\rm tot}\rangle^2}\sim\la\Bigl (\frac{m_Q\mph}{\la^2}\Bigr )^{(2N_c-N_F)/\nd}\,,\label{(3.7)}
\eq
\bq\frac{\mu^{\rm pole}_{3,S}(\Phi)}{\langle\lym^{(S)}\rangle}\sim\Bigl (\frac{m_Q\mph}{\la^2}\Bigr )^{2(3N_c-2N_F)/3\nd}\,<\,\Bigl (\frac{m_Q\mo}{\la^2}\Bigr )^{2(3N_c-2N_F)3\nd}\sim\Bigl (\frac{m_Q}{\la}\Bigr )^{2(3N_c-2N_F)/3N_c}\ll 1\,.\label{(3.8)}
\eq

The overall mass hierarchies look as
\bq
\mu^{\rm pole}_{3,S}(\Phi)\ll\langle\lym^{(S)}\rangle\ll m^{\rm pole}_{Q,S}\ll\mu^{\rm pole}_{2,S}(\Phi)\sim\mu_o^{\rm str}\ll\la\ll\mu_1^{\rm pole}(\Phi)\sim\mph\,.\label{(3.9)}
\eq

\subsection{\quad Dual theory}

The mass spectra in the Seiberg dual $SU(\nd)$ theory for this case were also described in section 7.2 of \cite{ch6}. The results look as follows. \\
1) All $N_F^2$ fions $\Phi^j_i$ have large masses $\mu_1^{\rm pole}(\Phi)\sim\mph\gg\la$ (with logarithmic accuracy) and are dynamically irrelevant at all lower scales.\\
2) All dual quarks ${\ov q}^j, q_i$ are in {\bf the overall Hq} (heavy quark) phase. There is a large number of hadrons made of weakly interacting non-relativistic and weakly confined dual quarks, the scale of their masses (with logarithmic accuracy) is $\mu^{\rm pole}_{q,S}\sim \langle M\rangle_{\rm S}/\la=\qq_{\rm S}/\la\sim (m_Q\mph)/\la\sim m^{\rm pole}_{Q,S}$, see \eqref{(2.3)},\eqref{(3.1)}\,  (the tension of the confining string originating from ${\cal N}=1\,\, SU(\nd)$ SYM is much smaller, $\sqrt \sigma\sim\langle\lym^{(S)}\rangle\ll\mu^{\rm pole}_{q,S}\sim m^{\rm pole}_{Q,S}$, see \eqref{(3.5)}\,).\\
3) A large number of gluonia made of $SU(\nd)$ gluons with their mass scale $\sim\langle\lym^{(S)}\rangle=\langle S\rangle_S^{1/3}\sim \la (m_Q\mph/\la^2)^{N_F/3\nd}$.\\
4) $N_F^2$ Seiberg's mions $M^i_j$ with masses $\mu^{\rm pole}_{S}(M)\sim\la^2/\mph$.\\

The hierarchies of masses (except for $\mu^{\rm pole}_{1}(\Phi)\sim\mph\gg\la$) look here as:\\
a) $\la\gg\mu^{\rm pole}_S(M)\gg\mu^{\rm pole}_q\gg\lym^{(\rm S)}$\quad at \quad $\la\ll\mph\ll{\mu_{\Phi}^\prime}=\la (\la/m_Q)^{1/2}\,$;\\
b) $\la\gg\mu^{\rm pole}_q\gg\mu^{\rm pole}_S(M)\gg\lym^{(\rm S)}$\quad at \quad ${\mu_{\Phi}^\prime}\ll\mph\ll {\tilde\mu}_{\Phi}=\la (\la/m_Q)^{N_F/(4N_F-3N_c)}\,$;\\
c) $\la\gg\mu^{\rm pole}_q\gg\lym^{(\rm S)}\gg\mu^{\rm pole}_S(M)$\quad at \quad ${\tilde\mu}_{\Phi}\ll\mph\ll\mo=\la (\la/m_Q)^{(2N_c-N_F)/N_c}\,$\,.

It is seen that, at least, $\mu^{\rm pole}_{3,S}(\Phi)\sim\la\Bigl (m_Q\mph/\la^2\Bigr )^{(2N_c-N_F)/\nd}$ \eqref{(3.7)} in the direct theory and $\mu^{\rm pole}_S(M)\sim\la^2/\mph$ in the dual one are parametrically different (the 't Hooft triangles $SU^3(N_F)_L$ are also different).

\section{Broken flavor symmetry, \, L-type vacua}

The main qualitative difference compared with the L-vacua in section 2 is the spontaneous breaking of the flavor symmetry, $U(N_F)\ra U(n_1)\times U(n_2)$, see Appendix,
\bbq
\langle ({\ov Q}Q)_1\rangle_{Lt}\equiv\langle {\ov Q}_1 Q^1\rangle_{Lt}=\langle M_1\rangle_{Lt}\neq \langle ({\ov Q}Q)_2\rangle_{Lt}=\langle {\ov Q}_2 Q^2\rangle_{Lt}=\langle M_2\rangle_{Lt}\,,
\eeq
\bq
(1-\frac{\no}{N_c})\langle ({\ov Q}Q)_1\rangle_{Lt}\simeq -\, (1-\frac{\nt}{N_c})\langle ({\ov Q}Q)_2\rangle_{Lt}\sim\qq_L\sim \la^2\Bigl (\frac{\la}{\mph}\Bigr )^{\nd/(\bb)}\ll\la^2\,.\label{(4.1)}
\eq

For this reason, unlike the L-vacua, the fions $\Phi^1_2,\, \Phi^2_1$ in the direct theory and mions $M^1_2,\, M^2_1$ in the dual one are the Nambu-Goldstone particles and are massless. Except for this, all other masses in these Lt-vacua of the direct and dual theories are parametrically the same as in L-vacua. Therefore, all differences between the direct and dual theories described above for the L-vacua remain in L-type vacua also.

\section{Broken flavor symmetry, \, br2 vacua}

\subsection{\quad Direct theory}

The quark condensates of the $SU(N_c$ theory look in these vacua with $\nt>N_c,\, \no<\nd$ as, see Appendix,
\bq
\langle\Qt\rangle_{\rm br2}\simeq\frac{N_c}{N_c-\nt} m_Q\mph\,,\quad \langle\Qo\rangle_{\rm br2}\sim\la^2
\Bigl (\frac{m_Q}{\la}\Bigr )^{\frac{N_c-n_1}{n_2-N_c}}\Bigl (\frac{\mph}{\la}\Bigr )^{\frac{n_2}{n_2-N_c}}\,,\label{(5.1)}
\eq
\bbq
\frac{\langle\Qo\rangle_{\rm br2}}{\langle\Qt\rangle_{\rm br2}}\sim\Bigl (\frac{\mph}{\mo}\Bigr )^{\frac{N_c}{n_2-N_c}}\ll 1\,,\quad \mo=\la\Bigl(\frac{\la}{m_Q}\Bigr )^{(2N_c-N_F)/N_c}\,,
\eeq
\bbq
\langle\lym^{(\rm br2)}\rangle^3\equiv\langle S\rangle_{\rm br2}=\Bigl (\frac{\langle\Qo\rangle_{\rm br2}^{\no}\langle\Qt\rangle_{\rm br2}^{\nt}}{\la^{3N_c-N_F}}\Bigr )^{1/\nd}
\sim\la^3\Bigl (\frac{m_Q}{\la}\Bigr )^{\frac{n_2-n_1}{n_2-N_c}}\Bigl (\frac{\mph}{\la}\Bigr )^{\frac{n_2}{n_2-N_c}}\,.
\eeq
From \eqref{(5.1)} and the Konishi anomalies,
\bbq
\qma_{\rm br2}=m_Q-\langle\Phi_1\rangle_{\rm br2}=\frac{\langle\Qt\rangle_{\rm br2}}{\mph}\sim m_Q\gg \qmb_{\rm br2}=\frac{\langle\Qo\rangle_{\rm br2}}{\mph}\sim\Bigl(\frac{m_Q}{\la}\Bigr )^{\frac{N_c-n_1}{n_2-N_c}}\Bigl (\frac{\mph}{\la}\Bigr )^{\frac{N_c}{n_2-N_c}},
\eeq
\bq
m_{Q,1}^{\rm pole}=\frac{\qma_{\rm br2}}{z_Q^{+}(\la,m_{Q,1}^{\rm pole})}\sim\la\Bigl (\frac{m_Q}{\la}\Bigr )^{\nd/N_c},\,\, z_Q^{+}(\la,m_{Q,1}^{\rm pole})=\Bigl (\frac{m_{Q,1}^{\rm pole}}{\la}\Bigr )^{\gamma_Q^{+}},\,\, \gamma_Q^{+}=\frac{2N_c-N_F}{N_F-N_c}>\, 1\,,\label{(5.2)}
\eq
while the gluon mass from the possible higgsing of ${\ov Q}_2, Q^2$ quarks looks here as, see \eqref{(2.4)},
\bbq
\Bigl ({\mu}^{\rm pole}_{gl,2}\Bigr )^2\sim\Biggl [a_{+}(\mu={\mu}_{gl,2})=\Bigl (\frac{\la}{{\mu}^{\rm pole}_{gl,2}}\Bigr )^{\nu^{+}_Q}\Biggr ] z_Q^{+}(\la,{\mu}^{\rm pole}_{gl,2})\langle\Qt\rangle_{\rm br2}\,, \quad \nu^{+}_Q=\gamma^{+}_Q-1=\frac{3N_c-2N_F}{N_F-N_c}>\,0\,,
\eeq
\bq
a_{+}({\mu}_{gl,2})\,z_Q^{+}(\la,{\mu}^{\rm pole}_{gl,2})\sim\frac{{\mu}^{\rm pole}_{gl,2}}{\la}\,,\quad
{\mu}^{\rm pole}_{gl,2}\sim\frac{\langle\Qt\rangle_{\rm br2}}{\la}\sim\frac{m_Q\mph}{\la}
\gg\mu_{gl,1}\,,\quad\frac{{\mu}^{\rm pole}_{gl,2}}{m_{Q,1}^{\rm pole}}\sim\frac{\mph}{\mo}\ll 1\,.\label{(5.3)}
\eq

Besides, see \eqref{(2.7)}, because the largest mass in the quark-gluon sector is $m_{Q,1}^{\rm pole}$ and
\bq
\frac{m_{Q,1}^{\rm pole}}{\mos}\sim\Bigl (\frac{m_Q}{\la}\Bigr )^{\frac{\nd}{N_c}} \Bigl (\frac{\mph}{\la}\Bigr )^{\frac{\nd}{5N_c-3N_F}}\ll\Bigl (\frac{m_Q}{\la}\Bigr )^{\frac{\nd}{N_c}} \Bigl (\frac{\mo}{\la}\Bigr )^{\frac{\nd}{5N_c-3N_F}}\sim\Bigl (\frac{m_Q}{\la}\Bigr )^{\frac{\nd(3N_c-2N_F)}{N_c(5N_c-3N_F)}\,>\,0}\ll 1\,,\label{(5.4)}
\eq
\bbq
\mos=\la\Bigl (\frac{\la}{\mph}\Bigr )^{\frac{1}{2\gamma^{+}_Q-1}}=\la\Bigl (\frac{\la}{\mph}\Bigr )^{\frac{\nd}{5N_c-3N_F}}\ll\la\,,\quad \nd=N_F-N_c\,,
\eeq
there are $N_F^2$ of 2-nd generation fions $\Phi^j_i$ with masses $\mu_2^{\rm pole}(\Phi)\sim\mos=\la(\la/\mph)^{\nd/(5N_c-3N_F)}\gg m_{Q,1}^{\rm pole}$.

After the heaviest quarks ${\ov Q}_1, Q^1$ decouple at $\mu<m_{Q,1}^{\rm pole}$, the RG evolution of $\Phi^1_1, \Phi^2_1, \Phi^1_2$ fions is frozen, while the new quantum numbers are
\bq
N_F^\prime=N_F-n_1=n_2\,,\,\,\, N_c^\prime=N_c\,,\,\, \gamma_Q^{-}=\frac{2N_c-N_F+n_1}{n_2-N_c}>
\gamma_Q^{+}\,,\,\,\, \nu^{-}_Q=\frac{3N_c-2N_F+2n_1}{n_2-N_c}>\nu^{+}_Q\,.\label{(5.5)}
\eq
From \eqref{(5.5)}, the pole mass of ${\ov Q}_2, Q^2$ quarks is, see \eqref{(5.2)},
\bq
m_{Q,2}^{\rm pole}=\frac{\qmb_{\rm br2}}{z_Q^{+}(\la,m_{Q,1}^{\rm pole}) z_Q^{-}(m_{Q,1}^{\rm pole},m_{Q,2}^{\rm pole})}\sim\frac{m_Q\mph}{\la}\,,\label{(5.6)}
\eq
\bbq
z_Q^{-}(m_{Q,1}^{\rm pole},m_{Q,2}^{\rm pole})=
\Bigl (\frac{m_{Q,2}^{\rm pole}}{m_{Q,1}^{\rm pole}}\Bigr )^{\gamma_Q^{-}}\,,\quad
\frac{m_{Q,2}^{\rm pole}}{m_{Q,1}^{\rm pole}}\sim\frac{\mph}{\mo}\ll 1\,.
\eeq
Besides, because $n_2>N_c$ in these br2-vacua, the quarks ${\ov Q}_2, Q^2$ are not higgsed in these vacua due to the rank restriction, as otherwise the flavor symmetry $U(n_2)$ will be further broken spontaneously. Therefore, {\bf the overall phase is $\mathbf {HQ_1-HQ_2}$}. All quarks are not higgsed but weakly confined, the confinement originates from ${\cal N}=1\,\, SU(N_c)$ \,SYM, so that the scale of the confining string is $\sqrt\sigma\sim\langle\lym^{(\rm br2)}
\rangle\ll m_{Q,2}^{\rm pole}$, see \eqref{(5.1)},\eqref{(5.6)}.

After the quarks ${\ov Q}_2, Q^2$ decouple at $\mu<m_{Q,2}^{\rm pole}$, there remain ${\cal N}=1 \,\, SU(N_c)$ SYM and $N_F^2$ fions $\Phi^j_i$. The value of the scale factor $\langle\lym^{\rm (br2)}\rangle$ of SYM is determined from the matching
\bbq
a_{-}(\mu=m_{Q,2}^{\rm pole})=\Bigl (\frac{\la}{m_{Q,1}^{\rm pole}}\Bigr )^{\nu^{+}_Q}\Bigl (\frac{m_{Q,1}^{\rm pole}}{m_{Q,2}^{\rm pole}}\Bigr )^{\nu^{-}_Q}=a^{\rm str}_{YM}(\mu=m_{Q,2}^{\rm pole})=\Bigl (\frac{m_{Q,2}^{\rm pole}}{\langle\lym^{\rm (br2)}\rangle}\Bigr )^3\quad \ra
\eeq
\bq
\quad \ra\quad \Bigl (\langle\lym^{\rm (br2)}\rangle\Bigr )^3\sim\la^3\Bigl(\frac{m_Q}{\la}\Bigr )^
{\frac{n_2-n_1}{n_2-N_c}}\Bigl (\frac{\mph}{\la}\Bigr )^{\frac{n_2}{n_2-N_c}}\,,\label{(5.7)}
\eq
as it should be, see \eqref{(5.1)},
\bq
\frac{\langle\lym^{\rm (br2)}\rangle}{m_{Q,2}^{\rm pole}}\sim\Bigl (\frac{m_Q}{\la}\Bigr )^{\frac{3N_c-2N_F+n_1}{3(n_2-N_c)}}\Bigl (\frac{\mph}{\la}\Bigr )^{\frac{3N_c-2N_F+2n_1}{3(n_2-N_c)}}\ll\Bigl (\frac{m_Q}{\la}\Bigr )^{\frac{3N_c-2N_F+n_1}{3(n_2-N_c)}}\Bigl (\frac{\mo}{\la}\Bigr )
^{\frac{3N_c-2N_F+2n_1}{3(n_2-N_c)}}\sim \label{(5.8)}
\eq
\bbq
\sim\Bigl (\frac{m_Q}{\la}\Bigr )^{\frac{(3N_c-2N_F)(\nd-n_1)}{3N_c(n_2-N_c)}\,>\,0}\ll 1\,,\quad n_1<\nd\,,\quad n_2>N_c\,.
\eeq

After integrating out all gluons at $\mu<\langle\lym^{\rm (br2)}\rangle$ via the VY procedure \cite{VY}, the lower energy Lagrangian of $N_F^2$ fions $\Phi^j_i$ looks as, see \eqref{(1.1)} for ${\cal W}_{\Phi}$ and \eqref{(5.3)},\eqref{(5.6)},
\bq
K_{\Phi}=z_{\Phi}^{+}(\la,m_{Q,1}^{\rm pole})\, {\rm Tr}\,\Bigl [ (\Phi^1_1)^\dagger \Phi^1_1+\Bigl ((\Phi^1_2)^\dagger \Phi^1_2+(\Phi^2_1)^\dagger \Phi^2_1)+z_{\Phi}^{-}(m_{Q,1}^{\rm pole},m_{Q,2}^{\rm pole})\, (\Phi^2_2)^\dagger \Phi^2_2 \Bigr ],\label{(5.9)}
\eq
\bbq
{\cal W}={\cal W}_{\Phi}+{\cal W}_{\rm{non-pert}},\quad {\cal W}_{\rm{non-pert}}=N_c\Bigl (\la^{\rm \bo}\det m_Q^{\rm tot}\Bigr )^{1/N_c},\quad m_Q^{\rm tot}=(m_Q-\Phi)\,,
\eeq

From \eqref{(5.9)}, the masses of the third generation fions look as, see \eqref{(5.1)},\eqref{(5.2)},\eqref{(5.6)},
\bq
\mu^{\rm pole}_{3}(\Phi_1^1)\sim\frac{\mph}{z_{\Phi}^{+}(\la,m_{Q,1}^{\rm pole})}=\mph\Bigl (z_Q^{+}(\la,m_
{Q,1}^{\rm pole})\Bigr )^2\sim \mph\Bigl (\frac{m_Q}{\la}\Bigr )^{\frac{2(2N_c-N_F)}{N_c}}\,,\label{(5.10)}
\eq
\bbq
z_{\Phi}^{\pm}(\mu_1,\mu_2)=(\frac{\mu_2}{\mu_1})^{\gamma_{\Phi}^{\pm}\,=\,-2\gamma_Q^{\pm}}\,,\quad\frac{\mu^{\rm pole}_{3}(\Phi_1^1)}{m_{Q,1}^{\rm pole}}\ll\frac{\mu^{\rm pole}_{3}(\Phi_1^1)}{m_{Q,2}^{\rm pole}}\sim\Bigl (\frac{m_Q}{\la}\Bigr )^{\frac{3N_c-2N_F}{N_c}}\ll 1\,,
\eeq
\bq
\mu^{\rm pole}_{3}(\Phi_2^2)\sim\frac{\langle S\rangle_{\rm br2}}{\qmb_{\rm br2}^2}\,\frac{1}{z_{\Phi}^{+}(\la,m_{Q,1}^{\rm pole})z_{\Phi}^{-}(m_{Q,1}^{\rm pole},m_{Q,2}^{\rm pole})}\sim\frac{\langle\Qt\rangle_{\rm br2}}{\langle\Qo\rangle_{\rm br2}}\,
\frac{\mu^{\rm pole}_{3}(\Phi_1^1)}{z_{\Phi}^{-}(m_{Q,1}^{\rm pole},m_{Q,2}^{\rm pole})}\,,\label{(5.11)}
\eq
\bbq
\mu^{\rm pole}_{3}(\Phi_2^2)\sim\la\Bigl (\frac{m_Q}{\la}\Bigr )^{\frac{2N_c-N_F}{n_2-N_c}}\Bigl (\frac{\mph}{\la}\Bigr)^{\frac{2N_c-N_F+n_1}{n_2-N_c}},\quad
\quad \frac{\mu^{\rm pole}_{3}(\Phi_2^2)}{\langle\lym^{\rm (br2)}\rangle}\ll 1\,,
\eeq
(the main contribution to $\mu^{\rm pole}_{3}(\Phi_1^1)$ originates from the term ${\cal W}_{\Phi}\sim \mph{\rm Tr}\,(\Phi^2)$ in \eqref{(5.9)}, while the main contribution to $\mu^{\rm pole}_{3}(\Phi_2^2)$ originates from ${\cal W}_{\rm non-pert}$ in \eqref{(5.9)}\,).

The fions $\Phi_1^2$ and $\Phi^1_2$ are the Nambu-Goldstone particles and are massless.\\

On the whole for this case the mass spectrum looks as follows.\\
1) Among the masses smaller than $\la$, the largest are the masses of $N_F^2$ 2-nd generation fions,
$\mu^{\rm pole}_2(\Phi)\sim\mos\sim\la(\la/\mph)^{\nd/(5N_c-3N_F)}$.\\
2) Next are masses of ${\ov Q}_1, Q^1$ quarks, $m_{Q,1}^{\rm pole}\sim\la (m_Q/\la)^{\nd/N_c}$, see \eqref{(5.2)}.\\
3) The masses of ${\ov Q}_2, Q^2$ quarks are $m_{Q,2}^{\rm pole}\sim m_Q\mph/\la\ll m_{Q,1}^{\rm pole}$, see \eqref{(5.6)}.\\
4) There is a large number of $SU(N_c)$ SYM gluonia with the mass scale $\sim\langle\lym^{\rm (br2)}\rangle\ll m_{Q,2}^{\rm pole}$, see \eqref{(5.7)},\eqref{(5.8)}.\\
5) The 3-rd generation fions $\Phi_1^1$ have masses $\mu^{\rm pole}_{3}(\Phi_1^1)
\sim\mph (m_Q/\la)^{2(2N_c-N_F)/N_c}\ll m_{Q,1}^{\rm pole}$, see \eqref{(5.10)}.\\
6) The 3-rd generation fions $\Phi_2^2$ have masses \eqref{(5.11)},  $\mu^{\rm pole}_{3}(\Phi_2^2)\ll\langle\lym^{\rm (br2)}\rangle$.\\
7) The 3-rd generation fions $\Phi_1^2, \Phi_2^1$ are the massless Nambu-Goldstone particles.

The overall hierarchy of nonzero masses look as
\bbq
\mu^{\rm pole}_{3}(\Phi_2^2)\ll\langle\lym^{\rm (br2)}\rangle\ll m_{Q,2}^{\rm pole}\ll m_{Q,1}^{\rm pole}\ll\mu^{\rm pole}_{2}(\Phi)\ll\la\ll\mu_1^{\rm pole}(\Phi)\sim\mph,
\eeq
\bbq
\mu^{\rm pole}_{3}(\Phi_2^2)\ll\mu^{\rm pole}_{3}(\Phi_1^1)\ll m_{Q,2}^{\rm pole}\ll m_{Q,1}^{\rm pole}\,,\quad \la\ll\mph\ll\mo=\la(\la/ m_Q)^{(2N_c-N_F)/N_c}.
\eeq
\vspace*{1mm}

Let us point out also the following, see section 8.1 in \cite{ch6}. The dual $SU(N_c)$ theory in br1-vacua considered therein looks as (see (8.1.10) in \cite{ch6}\,)
\bbq
K={\rm Tr}\,(\Phi^\dagger\Phi)+\frac{1}{\lt^2}{\rm Tr}\,(\tp^\dagger \tp)+{\rm Tr}\,({\tq}^\dagger \tq+
{\otq}^\dagger {\otq}),\quad \tm=\frac{N_c}{\nd}\,m\,,\quad \wmu=-\mx\,,\quad m\ll\mx\ll\lm\,,
\eeq
\bbq
W_{\rm matter}={\cal W}_{\Phi}+ {\rm Tr}\,\,(\tm-\Phi)\tp\,-\frac{1}{\lt}{\rm Tr}\,(\,{\otq}\, \tp \tq \,)\,,
\quad {\cal W}_{\Phi}=\frac{\wmu}{2}\Bigl (\,{\rm Tr}\,(\Phi^2)-\frac{1}{N_c}({\rm Tr}\,\Phi)^2\Bigr )\,.
\eeq

Let us compare now the mass spectra of this theory with those of \eqref{(1.1)}, with the substitution (see (8.1.24) in \cite{ch6}, both these $SU(N_c)$ theories
with $N_c<N_F<3N_c/2$ quark flavors are strongly coupled at scales $\mu<\la$):
\bq
\la=\lt=(-\lm)\Biggl (\frac{\lm}{\mx}\Biggr )^{\frac{\nd}{3N_c-2N_F}},\quad m_Q= m\,\frac{\lt}{\mph}\,, \quad \mph=\frac{\lt^2}{\mx}\,,
\quad m\ll\mx\ll\lm\,. \label{(5.12)}
\eq
The particle masses smaller than $\mx$ look then as follows\,:\\
a) the masses $m^{\rm pole}_{Q,1}$ of ${\ov Q}_1, Q^1$ quarks, see \eqref{(5.2)}, compare with (8.1.11) in \cite{ch6},
\bbq
m^{\rm pole}_{Q,1}\sim\la \Bigl (\frac{m_Q}{\la}\Bigr )^{\frac{\nd}{N_c}}\sim\lm (\frac{m}{\lm})^{{\frac{\nd}{N_c}}}\sim \mu_{\tq,1}^{\rm pole}\,,
\hspace*{9cm} \label{(5.12.a)}
\eeq
b) the masses $m^{\rm pole}_{Q,2}$ of ${\ov Q}_2, Q^2$ quarks, see \eqref{(5.6)}, compare with (8.1.11) in \cite{ch6},
\bbq
m_{Q,2}^{\rm pole}\sim \frac{m_Q\mph}{\la}\sim m\sim \mu_{\tq,2}^{\rm pole}\,,\hspace*{11.5cm} \label{(5.12.b)}
\eeq
c) the mass scale of gluonia, see \eqref{(5.7)}, compare with (8.1.8) in \cite{ch6},
\bbq
\langle\lym^{\rm (br2)}\rangle^3\sim\la^3\Bigl(\frac{m_Q}{\la}\Bigr )^{\frac{n_2-n_1}{n_2-N_c}}\Bigl (\frac{\mph}{\la}\Bigr )^{\frac{n_2}{n_2-N_c}}\sim
\mx m^2\Bigl (\frac{m}{\lm}\Bigr)^{\frac{2N_c-N_F}{\nt-N_c}}\sim\langle\lym^{\rm (br1)}\rangle^3\,,\hspace*{3.5cm} \label{(5.12.c)}
\eeq
d) The masses of 3-rd generation fions $\Phi_1^1$, see \eqref{(5.10)}, compare with (8.1.16) in \cite{ch6},
\bbq
\mu^{\rm pole}_{3}(\Phi_1^1)\sim\mph \Bigl (\frac{m_Q}{\la}\Bigr )^{\frac{2(2N_c-N_F)}{N_c}}\sim\mx\Bigl (\frac{m}{\lm} \Bigr )^{\frac{2(2N_c-N_F)}{N_c}}\sim\mu^{\rm pole}({M}_1^1)\,,\hspace*{5cm} \label{(5.12.d)}
\eeq
e) The masses of 3-rd generation fions $\Phi_2^2$, see \eqref{(5.11)}, compare with (8.1.17) in \cite{ch6},
\bbq
\mu^{\rm pole}_{3}(\Phi_2^2)\sim\la\Bigl (\frac{m_Q}{\la}\Bigr )^{\frac{2N_c-N_F}{n_2-N_c}}\Bigl (\frac{\mph}{\la}\Bigr )^{\frac{2N_c-N_F+n_1}{n_2-N_c}}\sim\mx\Bigl (\frac{m}{\lm}\Bigr )^{\frac{2N_c-N_F}{n_2-N_c}}\sim\mu^{\rm pole}({M}_2^2)\,,\hspace*{2.8cm} \label{(5.12.e)}
\eeq
f) The hybrids $\Phi^1_2$ and $M^1_2$ are massless.\\

Therefore, with the choice of parameters as in \eqref{(5.12)} and with correspondences $Q\leftrightarrow\textsf{q}$ and $\Phi\leftrightarrow M$, the
spectra of masses smaller than $\mx$ coincide in the direct $SU(N_c)$ theory \eqref{(1.1)} (this is the direct theory (8.1.18) in \cite{ch6}\,) and in the dual $SU(N_c)$ theory (8.1.10) in \cite{ch6}.

\subsection{\quad Dual theory}

For the reader convenience and to make this article self-contained, we reproduce below in short the results from the section 7.3 in \cite{ch6} for the mass spectra of the weakly coupled dual theory \eqref{(1.2)} (with \eqref{(5.12)}, the spectrum of masses smaller than $\mx$ in the case "A"\, below is the same as in the theories (8.1.7) and (8.1.19) in \cite{ch6}). \\

{\bf A) The range} $\mathbf{\la\ll\mph\ll \la(\la/m_Q)^{1/2}}$\\

{\bf The overall phase is $\mathbf{Higgs_1-Hq_2}$} in this range due to higgsing of dual quarks $q_1,{\ov q}^1$, $SU(\nd)\ra SU(\nd-n_1)$. On the whole for this case the mass spectrum looks as follows.

1) The heaviest (among the masses $<\la$) are $N_F^2$ mions $M^i_j$ with masses $\mu^{\rm pole}(M)\sim \la^2/\mph\,$.

2) There are $n_1(2\nd-n_1)$ massive dual gluons (and their super-partners) with masses \\ ${\ov\mu}^{\,\rm pole}_{\rm gl,1}\sim\langle N_1\rangle^{1/2}\sim (m_Q\la)^{1/2}$.

3) There is a large number of hadrons made of weakly interacting non-relativistic and weakly confined dual quarks $q^{\,\prime}_2$ and ${\ov q}^{\,\prime,\, 2}$ with unbroken colors, the scale of their masses is $\mu^{\rm pole}_{q,2}\sim m_Q\mph/\la\sim m_{Q,2}^{\rm pole}$, see \eqref{(5.6)}, (the tension of the confining string is much smaller, $\sqrt \sigma\sim\langle\lym^{(\rm br2)}\rangle\ll\mu^{\rm pole}_{q,2}$, see \eqref{(5.8)}).

4) The masses of $n_1^2$ nions $N_1^1$ (dual pions) are also $\mu^{\rm pole}(N_1^1)\sim m_Q\mph/\la$.

5) There is a large number of $SU(\nd-n_1)$ SYM gluonia, the scale of their masses is \\ $\sim\langle\lym^{(\rm br2)}\rangle\sim (m_Q\langle M_1\rangle_{\rm br2})^{1/3}=(m_Q\langle\Qo\rangle_{\rm br2})^{1/3}$, see \eqref{(5.1)},\eqref{(5.7)}.

6) Finally, $2n_1 n_2$ Nambu-Goldstone hybrid nions $N_1^2, N_2^1$ are massless.

The overall hierarchy of nonzero masses looks as:
\bq
\langle\lym^{(\rm br2)}\rangle\ll\mu(N_1^1)\sim\mu^{\rm pole}_{q,2}\ll{\ov\mu}_{\rm gl,1}\ll\mu^{\rm pole}(M)\ll\la\,.\label{(5.13)}
\eq

{\bf B) The range} $\mathbf{\la(\la/m_Q)^{1/2}\ll\mph\ll \mo=\la(\la/m_Q)^{(2N_c-N_F)/N_c}}$\\

The largest mass have in this case $q_2, {\ov q}^{\,2}$ quarks, see \eqref{(5.1)}
\bq
\mu^{\rm pole}_{q,2}\sim\frac{\langle M_2\rangle_{\rm br2}}{\la}=\frac{\langle\Qt\rangle_{\rm br2}}{\la}\sim\frac{m_Q\mph}{\la}\,.\label{(5.14)}
\eq
After integrating them out at $\mu<\mu^{\rm pole}_{q,2}$, the new scale factor of the gauge coupling is
\bq
\Bigl (\Lambda^{\prime\prime}\Bigr )^{3\nd-n_1}=\la^{\bd}\Bigl (\frac{m_Q\mph}{\la}\Bigr )^{n_2},
\quad\Lambda^{\prime\prime}\ll \mu^{\rm pole}_{q,2}\,,\quad \bd=3\nd-N_F\,, \quad n_1<\nd\,.\label{(5.15)}
\eq

In the range $(\la^3/m_Q)^{1/2}\ll\mph\ll {\mu}_{\Phi}^{\prime\prime}=\la\Bigl (\la/m_Q\Bigr )^{(3N_c-N_F-n_1)/2{\rm n}_2}\ll\mo$ the hierarchies look as: ${\ov\mu}_{\rm gl,1}\gg\Lambda^{\prime\prime}\gg\mu^{\rm pole}_{q,1}$, and therefore the quarks $q_1, {\ov q}^{\,1}$ are higgsed in the weak coupling region at $\mu={\ov\mu}_{\rm gl,1}\gg\Lambda^{\prime\prime}$, {\bf the overall phase is also $\mathbf{Higgs_1-Hq_2}$}. The mass spectrum looks in this region as follows (as previously, all mass values are given below up to logarithmic factors). - \\
1) There are $2n_1n_2$ unconfined dual quarks $q_2, {\ov q}^{\,2}$ with broken colors, their masses (up to logarithmic factors) are $\mu^{\rm pole}_{q,2}\sim \langle M_2\rangle_{\rm br2}/\la\sim m_Q\mph/\la$.\\
2) The quarks $q_2^{\prime}, {\ov q}^{\,\prime,\,2}$ with unbroken $SU(\nd-n_1)$ dual colors are confined and there is a large number of dual hadrons made from these weakly coupled and weakly confined non-relativistic quarks, the scale of their masses is also $\mu_{H}\sim (m_Q\mph/\la)$ (the tension of the confining string originated from $SU(\nd-n_1)\,\,\, {\cal N}=1$ SYM is $\sqrt\sigma\sim\langle\lym^{(\rm br2)}\rangle\ll\mu^{\rm pole}_{q,2} $).\\
3) $n_1(2\nd-n_1)$ massive gluons (and their super-partners) due to higgsing $SU(\nd)\ra SU(\nd-n_1)$ by $q_1, {\ov q}^{\,1}$ quarks, $\,\,{\ov\mu}^{\,\rm pole}_{\rm gl,1}\sim (m_Q\la)^{1/2}\ll\mu^{\rm pole}_{q,2} $.\\
3) $n_1^2$ nions $N_1^1$ and  $n_1^2$ mions $M_1^1$ also have masses $\mu^{\rm pole}(N_1^1)\sim\mu^{\rm pole}(M_1^1)\sim (m_Q\la)^{1/2}$.\\
4) There is a large number of gluonia from $SU(\nd-n_1)\,\,\, {\cal N}=1$ SYM, the scale of their masses is
$\sim \langle\lym^{(\rm br2)}\rangle\sim (m_Q\langle M_1\rangle_{\rm br2})^{1/3}\sim\la (m_Q/\la)^{({\rm n_2-n_1})/3(\nd-{\rm n}_1)}(\mph/\la)^{{\rm n}_2/3(\nd-{\rm n}_1)}$.\\
5) $n_2^2$ mions $M_2^2$ have masses $\mu^{\rm pole}(M_2^2)\sim\la^2/\mph$.\\
6) $2n_1 n_2$ mions $M_1^2$ and $M_2^1$ are the Nambu-Goldstone particles and are massless, $\mu(M_1^2)=\mu(M_2^1)=0$.

The overall hierarchy of nonzero masses looks in this range $(\la^3/m_Q)^{1/2}<\mu<{\mu}_{\Phi}^{\prime\prime}$ as
\bq
\mu(M_2^2)\ll\langle\lym^{(\rm br2)}\rangle\ll{\ov\mu}_{\rm gl,1}\sim\mu(N_1^1)\sim\mu(M_1^1)\ll
\mu^{\rm pole}_{q,2}\ll\la\,.\label{(5.16)}
\eq

But the hierarchies look as $\mu^{\rm pole}_{q,1}\sim (\langle M_1\rangle_{\rm br2}/\la)\gg\Lambda^{\prime
\prime}\gg{\ov\mu}_{\rm gl,1}$ at $\mu_{\Phi}^{\prime\prime}\ll\mph\ll\mo$, the quarks $q_1, {\ov q}^{\,1}$ are then too heavy and not higgsed, and {\bf the overall phase is $\mathbf{Hq_2-Hq_1}$} (heavy quarks). The mass spectrum looks in this region as follows.\\
1) All quarks are confined and there is a large number of dual hadrons made from these weakly coupled and weakly confined non-relativistic quarks, the scale of their masses is $\mu^{\rm pole}_{q,2}\sim (m_Q\mph/\la)\gg\mu^{\rm pole}_{q,1}\sim (\langle M_1\rangle_{\rm br2}/\la)=(\langle\Qo\rangle_{\rm br2}/\la)$, see \eqref{(5.1)} (the tension of the confining string originated from $SU(\nd)\,\,\, {\cal N}=1$ SYM is $\sqrt\sigma\sim\langle\lym^{(\rm br2)}\rangle\ll\mu^{\rm pole}_{q,1}\ll\mu^{\rm pole}_{q,2})$.\\
2) A large number of gluonia from $SU(\nd)\,\,\, {\cal N}=1$ SYM, the scale of their masses is
$\sim \langle\lym^{(\rm br2)}\rangle\sim (m_Q\langle M_1\rangle_{\rm br2})^{1/3}\sim\la (m_Q/\la)^{({\rm n_2-n_1})/3(\nd-{\rm n}_1)}(\mph/\la)^{{\rm n}_2/3(\nd-{\rm n}_1)}$.\\
3) $n_1^2$ mions $M_1^1$ with masses $\mu^{\rm pole}(M_1^1)\sim \la^2\langle M_2\rangle/\mph\langle M_1\rangle$.\\
4) $n_2^2$ mions $M_2^2$ with masses $\mu^{\rm pole}(M_2^2)\sim (\la^2/\mph)\ll\mu^{\rm pole}(M_1^1)$.\\
5) $2\no\nt$ mions $M_1^2$ and $M_2^1$ are the massless Nambu-Goldstone particles.

The overall hierarchy of nonzero masses look in this range $\mu_{\Phi}^{\prime\prime}<\mph<\mo$ as
\bq
\mu^{\rm pole}(M_2^2)\ll\mu^{\rm pole}(M_1^1)\ll\langle\lym^{(\rm br2)}\rangle\ll\mu^{\rm pole}_{q,1}\ll\mu^{\rm pole}_{q,2}\ll\la\,.\label{(5.17)}
\eq

Comparing with the direct theory in section 5.1 it is seen that the mass spectra are parametrically different.

\section{Broken flavor symmetry, special vacua}

\subsection{Direct theory}

The condensates in these vacua with $\no=\nd,\,\nt=N_c$ look as, see Appendix,
\bq
\langle\Qo\rangle=\frac{N_c}{2N_c-N_F} m_Q\mph,\quad\,\, \langle\Qt\rangle=\la^2\Bigl(\frac{\la}
{\mph}\Bigr )^{\frac{\nd}{2N_c-N_F}},\,\label{(6.1)}
\eq
\bbq
\langle S\rangle=\frac{\langle\Qo\rangle\langle\Qt\rangle}
{\mph}=\frac{N_c}{2N_c-N_F} m_Q\la^2\Bigl (\frac{\la}{\mph} \Bigr )^{\frac{\nd}{2N_c-N_F}}\,,\quad\frac{\langle\Qo
\rangle}{\langle\Qt\rangle}\sim\frac{m_Q}{\la}\Bigl (\frac{\mph}{\la}\Bigr )^{\frac{N_c}{2N_c-N_F}}\ll 1\,,
\eeq
\bbq
\langle m_{Q,1}^{\,\rm tot}\rangle=\frac{\langle\Qt\rangle}{\mph}=\la\Bigl (\frac{\la}{\mph}\Bigr )^{\frac{N_c}{2N_c-N_F}}\ll\la\,,\quad \langle m_{Q,2}^{\,\rm tot}\rangle=
\frac{\langle\Qo\rangle}{\mph}=\frac{N_c}{2N_c-N_F} m_Q\ll \langle m_{Q,1}^{\,\rm tot}\rangle\,.
\eeq

This direct theory is UV free and is in the strong coupling regime at $\mu<\la$, with the gauge coupling $a(\mu\ll\la)=(\la/\mu)^{\nu_Q\,>\,0}\gg 1$, see section 7 in \cite{ch1} and \eqref{(2.3)},\eqref{(2.4)}. The potentially most important masses look in these vacua as follows. The quark mass looks as, see \eqref{(6.1)},
\bq
m_{Q,1}^{\,\rm pole}=\frac{\langle m_{Q,1}^{\,\rm tot}\rangle}{z_Q(\la,m_{Q,1}^{\,\rm pole})}=\la\Bigl (\frac{\langle m_{Q,1}^{\,\rm tot}\rangle}{\la}\Bigr)^{{\frac{\nd}{N_c}}}=
\frac{\langle\Qt\rangle}{\la}\,, \label{(6.2)}
\eq
\bbq
z_Q(\la,\mu\ll\la)=\Bigl (\frac{\mu}{\la}\Bigr )^{\gamma_Q}\ll 1,\, \gamma_Q=\frac{2N_c-N_F}{N_F-N_c}>0\,.
\eeq
The gluon mass due to possible higgsing of quarks looks as, see \eqref{(6.1)},
\bq
\Bigl (\mu^{\,\rm pole}_{gl,2}\Bigr )^2\sim a(\mu=\mu^{\,\rm pole}_{gl,2})\,z_Q(\la,\mu^{\,\rm pole}_{gl,2})
\langle\Qt\rangle\sim\frac{\mu^{\,\rm pole}_{gl,2}}{\la}\,\langle\Qt\rangle\,,\label{(6.3)}
\eq
\bbq
a(\mu=\mu^{\,\rm pole}_{gl,2})\sim \Bigl (\frac{\la}{\mu^{\,\rm pole}_{gl,2}}\Bigr )^{\nu_Q}\,,\quad
z_Q(\la,\mu^{\,\rm pole}_{gl,2})=\Bigl (\frac{\mu^{\,\rm pole}_{gl,2}}{\la}\Bigr )^{\gamma_Q}\,,\quad \nu_Q=\gamma_Q-1=\frac{3N_c-2N_F}{N_F-N_c}>0,
\eeq
\bbq
m_{Q,2}^{\,\rm pole}\ll\mu^{\,\rm pole}_{gl,2}\sim\frac{\langle\Qt\rangle}{\la}\equiv\frac{\langle\Pi_2\rangle}{\la}\sim\la\Bigl (\frac{\la}{\mph} \Bigr )^{\frac{\nd}{2N_c-N_F}}\sim m_{Q,1}^{\,\rm pole}\ll\la\,,\quad\mu^{\,\rm pole}_{gl,2}\gg \mu^{\,\rm pole}_{gl,1}\,.
\eeq
{\bf The overall phase is $\mathbf{HQ_1-Higgs_2}$}. After integrating $Q^1,{\ov Q}_1$ quarks as heavy at $\mu\sim m_{Q,1}^{\,\rm pole}$, there remain $SU(N_c)$ SYM and $N^\prime_F=N_c$ quarks $Q^2,{\ov Q}_2$ which are higgsed at the same scale $\sim\langle\Qt\rangle/\la$. We use in this case $N^\prime_F=N_c$ the form of superpotential proposed in ~\cite{S1}.

As for the Kahler kinetic terms, we write them as
\bq
K={\rm Tr\,}\Biggl [z_{\Phi}(\la,m_{Q,1}^{\,\rm pole})\,\Phi^\dagger\Phi+
\frac{(\Pi^2_2)^{\dagger}\Pi^2_2}{\la^2}\,\Biggr ] +(B^{\dagger}_2 B_2+{\ov B}^{\,\dagger}_2\, {\ov B}_2)\,,\,\, \Pi^i_j=({\ov Q}_j Q^i)\,,\,\, i,j=\nd+1...N_F\,.\label{(6.4)}
\eq

The kinetic term $K_{\Pi}$ of $\nt^2=N_c^2$ pions $\Pi_2^2$ in \eqref{(6.4)} needs some explanations. There are two different contributions to $K_{\Pi}$. The first one originates directly from the kinetic term of higgsed quarks, see \eqref{(1.1)},\eqref{(6.2)},\eqref{(6.3)},
\bq
K_Q= z_Q^{+}(\la,\mgt){\,\rm Tr}\,\Biggl [\,(Q^{\,2}_2)^\dagger Q^2_2+ (Q^2_2\ra {\ov Q}^{\,2}_2)\,\Biggr ]\ra K^{\rm (Born)}_{\Pi}\sim z_Q^{+}(\la,\mgt)\,{\rm Tr}\,\sqrt{(\Pi^2_2)^\dagger\Pi^2_2}\,.\label{(6.5)}
\eq
The second one originates from the loop of either massive gluons or massive higgsed quarks (superpartners of massive gluons) integrated over the non-parametric interval of momenta $p_E\sim \mgt$. It looks parametrically as ( $(\mgt)^2\sim a_{+}(\mgt)z^{+}_Q(\la,\mgt)\sqrt{(\Pi^2_2)^\dagger\Pi^2_2}\,,\,\,$  $\,{\textsf Q}^2_2,\,{\ov {\textsf Q}}^{\,2}_2$ are canonically normalized quark fields ):
\bbq
K^{\rm (loop)}_{\Pi}\sim z_Q^{+}(\la,\mgt)\,{\rm Tr}\,\,\langle\langle\, (Q^{\,2}_2)^\dagger Q^2_2\,\rangle\rangle={\rm Tr}\,\,\langle\langle\, ({\textsf Q}^{\,2}_2)^\dagger {\textsf Q}^2_2\,\rangle\rangle\sim{\rm Tr}\,\int \frac{d^{\,4} p_E}{[\, p_{E}^2+(\mgt)^2\,]}\sim
\eeq
\bq
\sim {\rm Tr}\, (\mgt)^2\sim {\rm Tr}\,\Bigl [\, a_{+}(\mgt)z^{+}_Q(\la,\mgt)\sqrt{(\Pi^2_2)
^\dagger\Pi^2_2}\,\,\Bigr ]\,,\quad r=\frac{K^{\rm(loop)}_{\Pi}}{K^{\rm (Born)}_{\Pi}}\sim \,\, a_{+}(\mu=\mgt)\,.\label{(6.6)}
\eq
So, $r\ll 1$ if quarks were higgsed in the weak coupling regime $a_{+}(\mu=\mgt)\ll 1$, but $r\gg 1$ in our
case of the strong coupling regime $a_{+}(\mu=\mgt)\gg 1$. Therefore, $K_{\Pi}\sim K^{\rm (loop)}_{\Pi}$ is as in \eqref{(6.4)}.

As for the superpotential, it looks as, see \eqref{(1.1)} for ${\cal W}_{\Phi}$,
\bq
\w={\cal W}_{\Phi}+{\rm Tr\,} (m_{Q,2}^{\,\rm tot}\Pi^2_2)- {\rm Tr\,}\Bigl (\Phi^1_2\frac{\Pi^2_2}{m_{Q,1}^{\,\rm tot}}\Phi^2_1\Bigr )+{\cal W}_{\rm non-pert}\,,\label{(6.7)}
\eq
\bbq
{\cal W}_{\rm non-pert}=A\Biggl (1-\frac{\langle m_{Q,1}^{\,\rm tot}\rangle}{m_{Q,1}^{\,\rm tot}}\det\Bigl (\frac{\Pi^2_2}{\lambda^2}\Bigr )+\frac{{\ov B}_2 B_2}{\lambda^2}\Biggr )\,,\quad \lambda^2=\langle\Qt\rangle=\langle\Pi_2\rangle\,,\quad \langle A\rangle=\langle S\rangle \,.
\eeq
In \eqref{(6.7)}: $A$ is the Lagrange multiplier field, the additional factor $\langle m_{Q,1}^{\,\rm tot}\rangle/m_{Q,1}^{\,\rm tot}$ in ${\cal W}_{\rm non-pert}$ is needed to obtain from \eqref{(6.7)} the right value of $\langle\Phi_1\rangle$, the third term in $\w$ originated from integrating out heavy quarks $Q^{\,1},{\ov Q}_1$. Let us note that there is nothing "dual" in the form \eqref{(6.7)}. The $(\nt^2-1)$ independent light pion fields $\Pi^2_2$ originated in the standard way due to higgsed quarks $Q^{\,2},{\ov Q}_2$ (one degree of freedom $\delta\Pi^2_2\sim{\rm Tr} (\Pi^ 2_2-\langle\Pi^2_2\rangle)$ acquires the mass $\sim \mu^{\,\rm pole}_{gl,2}$ due to anomaly). The only non-trivial point is the appearance of two light baryons.The need for their appearance can be checked by counting numbers of degrees of freedom.

We obtain from \eqref{(6.4)},\eqref{(6.7)}:\\
a) the masses of $\nd^{\,2}$ third generation fions $\Phi^1_1$ are, see \eqref{(6.1)},\eqref{(6.2)},
\bq
\mu_{3}^{\rm pole}(\Phi^1_1)\sim\frac{\mph}{z_{\Phi}(\la,m_{Q,1}^{\,\rm pole})}\sim\frac{\la^2}{\mph}\,,\quad z_{\Phi}(\la,m_{Q,1}^{\,\rm pole})=\Bigl (\frac{m_{Q,1}^{\,\rm pole}}{\la}\Bigr )^{\gamma_{\Phi}}\sim\frac{\mph^2}{\la^2}\,,\quad \gamma_{\Phi}=-2\gamma_Q\,,\label{(6.8)}
\eq
(the main contribution to $\mu^{\rm pole}(\Phi^1_1)$ originates from the term ${\cal W}_{\Phi}$ in \eqref{(6.7)}, see \eqref{(1.1)}\,;\\

b) $N^2_c$ fields $\Pi^2_2$ and $N^2_c$ third generation fields $\Phi^2_2$ are mixed significantly and physical fields ${\widehat\Pi}^2_2$ and ${\widehat\Phi}^2_2$ have masses
\bq
\mu^{\rm pole}({\widehat\Pi}^{2}_2)\sim \mu_{3}^{\rm pole}({\widehat\Phi}^{2}_2)\sim\frac{\la^2}{\mph}\,;\label{(6.9)}
\eq
c) the baryons $B_2,\,{\ov B}_2$ have masses
\bq
\mu^{\rm pole}(B_2)=\mu^{\rm pole}({\ov B}_2)\sim\frac{\langle S\rangle}{\langle\Qt\rangle}=\langle m_{Q,2}^{\,\rm tot}\rangle\sim m_Q\,;\label{(6.10)}
\eq
d) $2\no\nt=2\nd N_c$ third generation hybrid fions $\Phi^1_2,\,\Phi^2_1$ are the Nambu-Goldstone particles and are massless.

The hierarchies of nonzero masses smaller than $\la$ look as
\bbq
\mu^{\rm pole}(B_2)\sim m_Q\ll\mu^{\rm pole}(\Phi^1_1)\sim\mu^{\rm pole}({\wh\Pi}^{2}_2)\sim \mu^{\rm pole}({\wh\Phi}^{2}_2)\sim\frac{\la^2}{\mph}\ll\mu^{\,\rm pole}_{gl,2}\sim m_{Q,1}^{\,\rm pole}\sim\la\Bigl (\frac{\la}{\mph} \Bigr )^{\frac{\nd}{2N_c-N_F}}\ll\la.
\eeq

\subsection{Dual theory}

The condensates look here as, see \eqref{(6.1)},
\bq
\langle M^i_j\rangle=\langle{\ov Q}_j Q^i\rangle,\quad \langle ({\ov q} q)_{i}\rangle=\frac{\langle S\rangle}{\langle ({\ov Q} Q)_i\rangle}\la=\langle m_{Q,i}^{\rm tot}\rangle\la,\quad i=1,\,2\,, \label{(6.11)}
\eq\bbq
\langle ({\ov q} q)_{1}\rangle=\frac{\langle ({\ov Q} Q)_2\rangle\la}{\mph}\,,\quad\langle ({\ov q} q)_{2}\rangle=\frac{\langle ({\ov Q} Q)_1\rangle\la}{\mph}\,.  \eeq

This dual theory is in the IR free logarithmic regime at $\mu<\la$ (all logarithmic factors of the RG evolution are ignored below for simplicity). All $N_F^2$ fions $\Phi^j_i$ with masses $\mu^{\rm pole}(\Phi)
\sim\mph\gg\la$ remain too heavy at scales $\mu<\la$ and can be integrated out from the beginning. The superpotential of the dual theory at $\mu=\la$ looks then as
\bq
{\cal W}={\cal W}_M-{\rm Tr}\,\Bigl ({\ov q}\,\frac{M}{\la}\, q \Bigr ),\quad
{\cal W}_M=m_Q{\rm Tr}\,M-\frac{1}{2\mph}\Bigl [\,{\rm Tr}\,(M^2)-\frac{1}{N_c}\Bigl ({\rm Tr}\, M\Bigr )^2\,\Bigr ]\,.\label{(6.12)}
\eq

The potentially most important masses look here as follows.\\
a) The pole masses of dual quarks
\bq
\mu^{\rm pole}_{q,2}\sim\mu_{q,2}\equiv\mu_{q,2}(\mu=\la)=\frac{\langle\Qt\rangle}{\la}\sim\la\Bigl (\frac{\la}{\mph}\Bigr )^{\frac{\nd}{2N_c-N_F}}\gg \mu^{\rm pole}_{q,1}\sim\frac{m_Q\mph}{\la}\,.\label{(6.13)}
\eq
b) The gluon masses due to possible higgsing of dual quarks
\bq
{\ov\mu}^{\,\rm pole}_{gl,1}\sim\langle ({\ov q} q)_1\rangle^{1/2}\sim\la\Bigl (\frac{\la}{\mph}\Bigr )^{\frac{N_c}{2(2N_c-N_F)}}\gg{\ov\mu}^{\,\rm pole}_{gl,2}\,,\quad
\Bigl (\frac{{\ov\mu}^{\,\rm pole}_{gl,1}}{\mu^{\rm pole}_{q,2}}\Bigr )^2\sim\Bigl (\frac{\la}{\mph}\Bigr )^
{\frac{3N_c-2N_F}{2N_c-N_F}}\ll 1\,,\label{(6.14)}
\eq
\bbq
\Bigl (\frac{{\ov\mu}^{\,\rm pole}_{gl,1}}{\mu^{\rm pole}_{q,1}}\Bigr )^2\sim\Bigl (\frac{\wh{\mu}_{\Phi}}{\mph}\Bigr )^{\frac{5N_c-2N_F}{2N_c-N_F}}\,,\quad \la\ll\wh{\mu}_{\Phi}=\la\Bigl (\frac{\la}{m_Q}\Bigr )^{\frac{2(2N_c-N_F)}{5N_c-2N_F}}\ll\mo\,.
\eeq

As it is seen from \eqref{(6.14)}, the hierarchy $\mu^{\rm pole}_{q,1} \lessgtr {\ov\mu}^{\,\rm pole}_{gl,1}$ changes at $\mph \lessgtr {\wh\mu}_{\Phi}$. As a result, {\bf the overall phase changes also}: it is  $\mathbf{Higgs_1-Hq_2}$ at  $\la\ll\mph\ll{\wh\mu}_{\Phi}$ and $\mathbf{Hq_1-Hq_2}$ at  ${\wh\mu}_{\Phi}\ll\mph\ll\mo$.\\

{\bf A)}.\,\,  Consider first the region $\la\ll\mph\ll {\wh\mu}_{\Phi}$. The dual quarks $q_2, {\ov q}^{\,2}$ decouple as heavy in the weak coupling regime at $\mu=\mu^{\rm pole}_{q,2}\ll\la$ and there remains $SU(\nd)$ with $\no=\nd$ flavors of quarks $q_1, {\ov q}^{\,1}$. The scale factor of the gauge coupling is
\bq
\Bigl (\wh\Lambda\Bigr )^{3\nd-\no=2\nd}=\la^{3\nd-N_F}\mu^{N_c}_{q,2}\,,\quad \wh\Lambda=\la\Bigl (\frac{\la}{\mph}\Bigr )^{\frac{N_c}{2(2N_c-N_F)}}\sim {\ov \mu}^{\,\rm pole}_{gl,1}.\label{(6.15)}
\eq

After integrating out $q_2, {\ov q}^{\,2}$ quarks as heavy at $\mu<\mu^{\rm pole}_{q,2}$, the dual Lagrangian looks as (all logarithmic factors in the Kahler term are ignored for simplicity)
\bbq
K={\rm Tr\,} \Bigl [\frac{M^{\dagger} M}{\la^2}+(q_1)^{\dagger} q_1 +(q_1\ra {\ov q}^{\,1}) \Bigr ]\,,
\eeq
\bq
\w={\cal W}_M + {\rm Tr}\,\Biggl ({\ov q}^{\,1} M^2_1\frac{1}{\la M^2_2} M^1_2 q_1\Biggr ) -{\rm Tr}\,\Bigl ({\ov q}^{\,1}\frac{M^1_1}{\la} q_1\Bigr )\,,\label{(6.16)}
\eq
${\cal W}_M$ is given in \eqref{(6.12)}.

All $\nd$ flavors of quarks $q_1, {\ov q}^{\,1}$ are higgsed at $\mu\sim {\ov \mu}^{\,\rm pole}_{gl,1}
\sim\wh\Lambda$, and we use for this case $N_F^\prime=\no=\nd$ the Seiberg form of ${\cal W}_{\rm non-pert}$ proposed in \cite{S1},
\bbq
K={\rm Tr\,} \Bigl [\frac{M^{\dagger} M}{\la^2}+ 2\sqrt{(N^1_1)^{\dagger} N^1_1}\,\Biggr ] +\Bigl (b^{\dagger}_1 b_1+ (b_1\ra {\ov b}_1)\,\Bigr )\,,\quad N^j_i=({\ov q}^{\,j} q_i)\,,\,\, i,j=1...\nd\,,
\eeq
\bq
\w={\cal W}_M - {\rm Tr\,}\,\frac{M^1_1 N^1_1}{\la} +{\rm Tr\,}\,\Biggl ( M^2_1\frac{N^1_1}{\la M^2_2} M^1_2\Biggr )+{\cal W}_{\rm non-pert}\,,\label{(6.17)}
\eq
\bq
{\cal W}_{\rm non-pert}={\ov A}\Biggl (1-\frac{\langle M^2_2\rangle}{M^2_2}\det\Bigl (\frac{N^1_1}{\langle N_1\rangle}\Bigr )+\frac{{\ov b}_1 b_1}{\langle N_1\rangle} \Biggr ),\,\, \langle{\ov A}\rangle=\langle {\ov S}\rangle=\langle -S\rangle,\,\,\langle N_1\rangle=\langle\qo\rangle\sim (\wh\Lambda)^2,\,\,\label{(6.18)}
\eq
where ${\ov A}$ is the (dual) Lagrange multiplier field, $N^1_1$ are dual pions (nions), the factor
$\langle M^2_2\rangle/M^2_2$ in ${\cal W}_{\rm non-pert}$ is needed to obtain the right value of $\langle M^2_2\rangle$, $\,\wh\Lambda$ is given in \eqref{(6.15)}.\\
We obtain from \eqref{(6.18)} for the particle masses\,:\\
a) fields $M^1_1$ and fields $N^1_1$  have masses
\bq
\mu^{\,\rm pole}( M^{\,1}_1)\sim\mu^{\,\rm pole}( N^{\,1}_1)\sim{\ov \mu}^{\,\rm pole}_{gl,1}\sim\la\Bigl (\frac{\la}{\mph}\Bigr )^{\frac{N_c}{2(2N_c-N_F)}}\,,\label{(6.19)}
\eq
b) the masses of baryons are
\bq
\mu^{\,\rm pole}(b_1)=\mu^{\,\rm pole}({\ov b}_1)\sim\frac{\langle M_1\rangle}{\la}\sim\frac{m_Q\mph}{\la}\ll\mu^{\,\rm pole}( M^{\,1}_1)\,,\quad
\frac{\mu^{\,\rm pole}(b_1)}{{\ov\mu}^{\,\rm pole}_{gl,1}}\sim \Bigl (\frac{\mph}{{\wh\mu}_{\Phi}}\Bigr )^{\frac{5N_c-2N_F}{2(2N_c-N_F)}}\ll 1\,,   \label{(6.20)}
\eq
c) $N_c^2$ mions $M^2_2$ have masses
\bq
\mu^{\,\rm pole}(M^2_2)\sim \frac{\la^2}{\mph}\ll\la\,,\label{(6.21)}
\eq
d) $2\no\nt=2\nd N_c$ hybrid mions $M^1_2,\, M^2_1$ are the Nambu-Goldstone particles and are massless.

The hierarchies of nonzero masses look in this region $\la\ll\mph\ll {\wh\mu}_{\Phi}=\la\Bigl (\frac{\la}{m_Q}\Bigr )^{\frac{2(2N_c-N_F)}{5N_c-2N_F}}$ as
\bq
\mu^{\,\rm pole}(M^2_2)\sim \frac{\la^2}{\mph}\ll\mu^{\,\rm pole}( M^{\,1}_1)\sim\mu^{\,\rm pole}( N^{\,1}_1)\sim{\ov \mu}^{\,\rm pole}_{gl,1}\sim\la\Bigl (\frac{\la}{\mph}\Bigr )^{\frac{N_c}{2(2N_c-N_F)}}\ll\quad \label{(6.22)}
\eq
\bbq
\ll\mu^{\rm pole}_{q,2}\sim\la\Bigl (\frac{\la}{\mph}\Bigr )^{\frac{\nd}{2N_c-N_F}}\ll\la\,,\quad \mu^{\,\rm pole}(M^2_2)\gg\mu^{\,\rm pole}(b_1)\sim\frac{m_Q\mph}{\la}\,.
\eeq

{\bf B)}\,\,  The region $\la\ll{\wh\mu}_{\Phi}\ll\mph\ll\mo=\la(\la/m_Q)^{(2N_c-N_F)/N_c}$. The superpotential \eqref{(6.12)} remains the same, the difference at lower energies is that quarks ${\ov q}^1, q_1$ are not higgsed now but decouple as heavy at $\mu<\mu^{\rm pole}_{q,1}\sim m_Q\mph/\la\ll\mu^{\rm pole}_{q,2}\ll\la$, still in the weak coupling regime, see \eqref{(6.13)},\eqref{(6.14)}. The scale factor of the gauge coupling of remained unbroken $SU(\nd)$ SYM at $\mu<\mu^{\rm pole}_{q,1}$ looks as, see \eqref{(6.13)},\eqref{(6.1)},
\bq
\lambda_{YM}^{3\nd}=\la^{3\nd-N_F}\mu_{q,2}^{N_c}\,\mu_{q,1}^{\nd}\,,\quad \lambda_{YM}^3=m_Q\la^2\Bigl (\frac{\la}{\mph}\Bigr )^{\frac{\nd}{2N_c-N_F}}=\langle\lym^{(\rm spec)}\rangle^3\,, \label{(6.23)}
\eq
as it should be. Therefore, after integrating out all quarks as heavy ones and then all $SU(\nd)$ gluons at $\mu=\langle\lym^{(\rm spec)}\rangle$ via the VY procedure \cite{VY}, the Lagrangian looks as, see \eqref{(6.12)} for ${\cal W}_M$,
\bq
K={\rm Tr\,}\frac{M^{\dagger} M}{\la^2}\,,\quad \w={\cal W}_M+{\cal W}_{\rm non-pert}\,,\quad {\cal W}_{\rm non-pert}=-\nd\Bigl (\la^{3\nd-N_F}\det\frac{M}{\la}\Bigr )^{1/\nd}\,.\label{(6.24)}
\eq
From \eqref{(6.24)}, the masses of $N^2_c$ mions $M^2_2$ and $\nd^2$ mions $M^1_1$ are (up to logarithmic factors), see \eqref{(6.1)},
\bq
\mu^{\,\rm pole}(M^2_2)\sim\frac{\la^2}{\mph}\ll\mu^{\,\rm pole}(M^1_1)\sim\frac{\langle\Qo\rangle}{\langle
\Qt\rangle}\frac{\la^2}{\mph}\sim\frac{\la^2}{m_Q}\Bigl (\frac{\la}{\mph}\Bigr )^{\frac{3N_c-N_F}{2N_c-N_F}}
\ll\langle\lym^{(\rm spec)}\rangle,\,\, {\wh\mu}_{\Phi}\ll\mph\ll\mo\,\,\, \label{(6.25)}
\eq
(the main contribution to $\mu^{\,\rm pole}(M^2_2)$ in \eqref{(6.24)} originates from ${\cal W}_M$, while $\mu^{\,\rm pole}(M^1_1)$ is dominated by the contribution from ${\cal W}_{\rm non-pert}$ ).

$2\nd N_c$ hybrid mions $M^1_2, M^2_1$ are the Numbu-Goldstone particles and are massless.

The hierarchies of nonzero masses look in this region ${\wh\mu}_{\Phi}=\la\Bigl (\frac{\la}{m_Q}\Bigr )^{\frac{2(2N_c-N_F)}{5N_c-2N_F}}\ll\mph\ll\mo$ as
\bq
\mu^{\,\rm pole}(M^2_2)\sim\frac{\la^2}{\mph}\ll\mu^{\,\rm pole}(M^1_1)\sim\frac{\la^2}{m_Q}\Bigl (\frac{\la}{\mph}\Bigr )^{\frac{3N_c-N_F}{2N_c-N_F}}
\ll\langle\lym^{(\rm spec)}\rangle\ll
\eq
\bbq
\ll\mu^{\rm pole}_{q,1}\sim\frac{m_Q\mph}{\la}\ll\mu^{\rm pole}_{q,2}\sim\la\Bigl (\frac{\la}{\mph}\Bigr )^{\frac{\nd}{(2N_c-N_F)}}\ll\la\,.
\eeq

\addcontentsline{toc}{section}
 {\hspace*{3cm} The region $\mo=\la (\la/m_Q)^{(2N_c-N_F)/N_c}\ll\mph\ll\la^2/m_Q$}
\vspace*{2mm}
\begin{center}{\Large\bf The region $\mo=\la (\la/m_Q)^{(2N_c-N_F)/N_c}\ll\mph\ll\la^2/m_Q$} \end{center}

\section{Unbroken flavor symmetry, \, QCD vacua}

\subsection{Direct theory.}

The direct theory is in the strong coupling regime at $\mu<\la$. The quark condensates look here as, see Appendix,
\bq
\qq_{\rm QCD}\simeq\la^2\Bigl (\frac{m_Q}{\la}\Bigr )^{\nd/N_c},\quad \langle\lym^{\rm (QCD)}\rangle ^3
\equiv\langle S\rangle_{\rm QCD}\simeq\Bigl (\la^{\rm \bo}m_Q^{N_F}\Bigr )^{1/N_c}\,,\quad \bo=3N_c-N_F\,.\label{(7.1)}
\eq
The potentially important masses look here as
\bbq
\langle m^{\rm tot}_Q\rangle_{\rm QCD}=\langle m_Q-\Phi\rangle_{\rm QCD}=\frac{\langle S\rangle_{\rm QCD}}{\qq_{\rm QCD}}\simeq m_Q\,,
\eeq
\bq
m^{\rm pole}_{Q,\rm QCD}=\frac{\langle m^{\rm tot}_Q\rangle_{\rm QCD}}{z_Q(\la,m^{\rm pole}_{Q,\rm QCD})}\sim \la\Bigl (\frac{m_Q}{\la}\Bigr )^{\nd/N_c},\,\, z_Q(\la,m^{\rm pole}_{Q,\rm QCD})=\Bigl (\frac{m^{\rm pole}_{Q,\rm QCD}}{\la}\Bigr )^{\gamma_Q=(2N_c-N_F)/\nd}\,,\label{(7.2)}
\eq
\bq
\mu^2_{\rm gl,QCD}\sim a_{+}(\mu=\mu_{\rm gl,QCD})z_Q(\la,\mu_{\rm gl,QCD})\qq_{\rm QCD}\quad\ra\quad
\mu_{\rm gl,QCD}\sim\frac{\qq_{\rm QCD}}{\la}\sim m^{\rm pole}_{Q,\rm QCD}\,,\label{(7.3)}
\eq
\bbq
a_{+}(\mu=\mu_{\rm gl,QCD})\sim \Bigl (\frac{\la}{\mu_{\rm gl,QCD}}\Bigr )^{\nu_Q=(3N_c-2N_F)/\nd},\quad
a_{+}(\mu=\mu_{\rm gl,QCD})\,z_Q(\la,\mu_{\rm gl,QCD})\sim\frac{\mu_{\rm gl,QCD}}{\la}\,.
\eeq
As before for vacua with the unbroken flavor symmetry, this implies from the rank restriction at $N_F>N_c$ that quarks are not higgsed but confined and {\bf the overall phase is HQ} (heavy quark).

Besides, see {\eqref{(5.4)} for $\mos$ and \eqref{(7.2)},
\bq
\frac{m^{\rm pole}_{Q,\rm QCD}}{\mos}\sim \Bigl (\frac{\mph}{{\hat \mu}_{\Phi,\rm QCD}}
\Bigr )^{\frac{\nd}{5N_c-3N_F}},\quad {\hat \mu}_{\Phi,\rm QCD}=\la\Bigl (\frac{\la}{m_Q}\Bigr )^{\frac{5N_c-3N_F}{N_c}}\gg\mo\,,\quad \frac{1}{2}<\frac{5N_c-3N_F}{N_c}<2\,.\label{(7.4)}
\eq

It follows from \eqref{(7.4)} that\,:\, a) $\,{m^{\rm pole}_{Q,\rm QCD}}<\mos$ in the range $\mo<\mph<{\hat \mu}_{\Phi,\rm QCD}$ and so there will be two additional generations of $\Phi$-particles, $\mu_{2}^{\rm pole}(\Phi)\sim\mos$ and $\mu_{3}^{\rm pole}(\Phi)\ll m^{\rm pole}_{Q,\rm QCD}$, and all $N_F^2$ fions $\Phi^j_i$ will be dynamically relevant at scales $\mu_{3}^{\rm pole}(\Phi)<\mu<\mu_{2}^{\rm pole}(\Phi)$;\, b)\, while ${m^{\rm pole}_{Q,\rm QCD}}>\mos$ at $\mph>{\hat \mu}_{\Phi,\rm QCD}$ and all fions will be too heavy and dynamically irrelevant at all scales $\mu<\mu_{1}^{\rm pole}(\Phi)\sim\mph$.

Proceeding further in a "standard" way, i.e. integrating out first all quarks as heavy ones at $\mu<m^{\rm pole}_{Q,\rm QCD}$, the scale factor of the lower energy $SU(N_c)$ SYM in the strong coupling regime is determined from the matching
\bbq
a_{+}(\mu=m^{\rm pole}_{Q,\rm QCD})=\Bigl (\frac{\la}{m^{\rm pole}_{Q,\rm QCD}}\Bigr )^{(3N_c-2N_F)/\nd}=a^{\rm str}_{\rm YM}(\mu=m^{\rm pole}_{Q,\rm QCD})=\Bigl (\frac{m^{\rm pole}_{Q,\rm QCD}}{\langle\lym^{(\rm QCD)}\rangle}\Bigr )^3\quad\ra
\eeq
\bq
\ra\quad\langle\lym^{(\rm QCD)}\rangle=\Bigl (\la^{3N_c-N_F}m_Q^{N_F}\Bigr )^{1/3N_c} \label{(7.5)}
\eq
as it should be, see \eqref{(7.1)}.

Integrating out now all $SU(N_c)$ gluons at $\mu<\lym^{(\rm QCD)}$ via the VY-procedure \cite{VY},
we obtain the low energy Lagrangian of $N_F^2$ fions $\Phi^j_i$
\bq
K=z_{\Phi}(\la,m^{\rm pole}_{Q,\rm QCD})\,{\rm Tr}\,(\Phi^\dagger\Phi)\,,\quad z_{\Phi}(\la,m^{\rm pole}_{Q,\rm QCD})=\Bigl(\frac{m^{\rm pole}_Q}{\la}\Bigr )^{\gamma_{\Phi}}=
\Bigl(\frac{\la}{m^{\rm pole}_Q}\Bigr )^{2(\bb)/\nd}\gg 1\,,\label{(7.6)}
\eq
\bbq
{\cal W}={\cal W}_{\Phi}+N_c\Bigl(\la^{\rm \bo}\det m^{\rm tot}_Q\Bigr )^{1/N_c}\,,\quad m^{\rm tot}_Q=m_Q-\Phi\,,\quad{\cal W}_{\Phi}=\frac{\mph}{2}\Biggl [{\rm Tr}\,(\Phi^2)-\frac{1}{\nd}\Bigl ({\rm Tr}\,\Phi\Bigr )^2\Biggr ].
\eeq
The main contribution to the masses of $N_F^2$ 3-rd generation fions at $\mo<\mph<{\hat\mu}_{\Phi,\rm QCD}$ originates from the term ${\cal W}_{\Phi}$ in \eqref{(7.6)}
\bq
\mu_{3}^{\rm pole}(\Phi)\sim\frac{\mph}{z_{\Phi}(\la,m^{\rm pole}_{Q,\rm QCD})}
\sim\mph\Bigl(\frac{m_Q}{\la}\Bigr )^{2(\bb)/N_c}\,,\label{(7.7)} \label{(7.7)}
\eq
\bbq
\quad \frac{\mu_{3}^{\rm pole}(\Phi)}{m^{\rm pole}_{Q,\rm QCD}} \sim\frac{\mph}{\la}
\Bigl(\frac{m_Q}{\la}\Bigr )^{\frac{5N_c-3N_F}{3N_c}}=\frac{\mph}{{\hat\mu}_{\Phi,\rm QCD}}<1\,.
\eeq

On the whole for the mass spectra in these $N_c$ QCD-vacua at $\mph>\mo=\la(\la/m_Q)^{(\bb)/N_c}$. {\bf The overall phase is HQ}, all quarks ${\ov Q}_j, Q^i$ are not higgsed but confined.

{\bf A}) The range $\mo=\la(\la/m_Q)^{(\bb)/N_c}<\mph<{\hat\mu}_{\Phi,\rm QCD}=\la (\la/m_Q)^{(5N_c-3N_F)/N_c}$.\\
1) The largest among masses $<\la$ are masses of $N_F^2$ second generation fions $\Phi^j_i$ with $\mu_2^{\rm pole}(\Phi)\sim\mos=\la(\la/\mph)^{\nd/(5N_c-3N_F)}$.\\
2) The next are masses $m^{\rm pole}_{Q,\rm QCD}\sim \la (m_Q/\la)^{\nd/N_c}$ of strongly interacting but weakly confined quarks (the tension of the confining string originating from ${\cal N}=1\,\, SU(N_c)$ SYM is $\sqrt\sigma\sim\lym^{(\rm QCD)}\ll m^{\rm pole}_{Q,\rm QCD}$).\\
3) There are $N_F^2$ third generation fions $\Phi^j_i$ with $\mu_3^{\rm pole}(\Phi)\sim\mph (m_Q/\la)^{2(\bb)/N_c}$.
\\
4) A large number of gluonia from ${\cal N}=1\,\, SU(N_c)$ SYM, the scale of their masses is $\lym^{(\rm QCD)}\sim (\la^{\rm \bo}m_Q^{N_F})^{1/3N_c}$.

The mass hierarchies look as
\bq
\mu_3^{\rm pole}(\Phi)\ll m^{\rm pole}_{Q,\rm QCD}\ll\mu_2^{\rm pole}(\Phi)=\mos\ll\la\ll\mu_1^{\rm pole}(\Phi)\sim\mph\,,\quad \lym^{(\rm QCD)}\ll  m^{\rm pole}_{Q,\rm QCD}\,,\label{(7.8)}
\eq
\bbq
\frac{\mu_3^{\rm pole}(\Phi)}{\lym^{(\rm QCD)}}=\left\{
\begin{array}{rl} < 1 & {\rm at}\quad \mo<\mph<{\tilde\mu}_{\Phi,\rm QCD}=
\la (\la/m_Q)^{(12N_c-7N_F)/3N_c} \\
> 1 &  {\rm at}\quad {\tilde\mu}_{\Phi,\rm QCD}<\mph<{\hat\mu}_{\Phi,\rm QCD}\,. \end{array}\right.   \\
\eeq

{\bf B})  The range $\mph>{\hat\mu}_{\Phi,\rm QCD}=\la (\la/m_Q)^{(5N_c-3N_F)/N_c}\gg{\tilde\mu}_{\Phi,\rm QCD}\gg\mo$\,.\\
The difference is that there are no additional generations of all fions in this case because $m^{\rm pole}_{Q,\rm QCD}>\mos$, and all fions remain too heavy and dynamically irrelevant at all scales $\mu<\mu_1^{\rm pole}(\Phi)\sim\mph$.

\subsection{Dual theory.}

The dual theory is in the IR free logarithmic regime at scales $\mu_{q,QCD}^{\rm pole}<\mu<\la$ (as before, all logarithmic factors are ignored for simplicity). The potentially important masses look
here as follows. The masses of dual quarks
\bq
\mu_{q,QCD}^{\rm pole}\sim\frac{\langle M\rangle_{\rm QCD}}{\la}=\frac{\langle {\ov Q}Q\rangle_{\rm QCD}}{\la}\sim\la\Bigl(\frac{m_Q}{\la}\Bigr )^{\frac{\nd}{N_c}}\sim m^{\rm pole}_{Q,\rm QCD}.\label{(7.9)}
\eq
The masses of dual gluons due to possible higgsing of dual quarks
\bq
{\ov\mu}^{\,\rm pole}_{\rm gl,QCD}\sim \langle ({\ov q}q)\rangle^{1/2}_{\rm QCD}\sim (m_Q\la)^{1/2},\quad
\frac{{\ov\mu}^{\,\rm pole}_{\rm gl,QCD}}{\mu_{q,QCD}^{\rm pole}}\sim\Bigl (\frac{m_Q}{\la}\Bigr )^{\frac{3N_c-2N_F}{2N_c}}\ll 1\,.\label{(7.10)}
\eq
Therefore, {\bf the overall phase is Hq} (heavy quark), all quarks ${\ov q}^j, q_i$ are not higgsed but confined. All $N_F^2$ fions $\Phi^j_i$ have large masses $\mu^{\rm pole}_{1}(\Phi)\sim\mph\gg\la$ and are irrelevant at scales $\mu<\mph$, they can be integrated out from the beginning. After integrating out all dual quarks as heavy ones at $\mu<\mu_{q,QCD}^{\rm pole}$ and then all dual gluons at $\mu<\langle\lym^{(QCD)}\rangle$ via the VY procedure \cite{VY}, the low energy Lagrangian of $N_F^2$ mions $M^i_j$ looks as
\bq
K\sim \,{\rm Tr}\,\frac{M^\dagger M}{\la^2},\,\, {\cal W}={\cal W}_M-\nd\Bigl (\frac{\det M}{\la^{\rm \bo}}\Bigr )^{1/\nd},\,\, {\cal W}_M=m_Q {\rm Tr}\,M-\frac{1}{2\mph}\Bigl [{\rm Tr}\,(M^2)-\frac{1}{N_c}\Bigl ({\rm Tr}\, M\Bigr )^2\Bigr ].\,\,\,\, \label{(7.11)}
\eq
The main contribution to the masses of $N_F^2$ mions originates from the nonperturbative term in \eqref{(7.11)}
\bq
\mu^{\rm pole}(M)\sim\frac{\langle S\rangle_{\rm QCD}\la^2}{\langle M\rangle^2_{\rm QCD}}\sim\la\Bigl(\frac{m_Q}{\la}\Bigr )^{(\bb)/N_c}\,,\quad \frac{\mu^{\rm pole}(M)}{\lym^{(QCD)}}\sim\Bigl (\frac{m_Q}{\la}\Bigr )^{\frac{2(3N_c-2N_F)}{3N_c}}\ll 1\,.\label{(7.12)}
\eq

Comparing with the direct theory in section 7.1 it is seen that at $\mo<\mph<{\hat\mu}_{\Phi,\rm QCD}$ the masses $\mu_3^{\rm pole}(\Phi)$ \eqref{(7.7)} and $\mu^{\rm pole}(M)$ \eqref{(7.12)} are parametrically different (the triangles $SU^3(N_F)_L$ are also different). Besides, there are no particles with masses smaller than $\langle\lym^{(QCD)}\rangle$ at $\mph>{\hat\mu}_{\Phi,\rm QCD}$ in the direct theory while $\mu^{\rm pole}(M)\ll\langle\lym^{(QCD)}\rangle$ in the dual one, see \eqref{(7.12)}.

\section{Broken flavor symmetry.\, Direct theory.}

\subsection{\quad  br1 and special vacua with $1\leq n_1<{\rm Min}\,(N_F/2,\,3N_c-2N_F)$}

The condensates look in this case as, see Appendix,
\bq
\langle\Qo\rangle_{\rm br1}\simeq\frac{N_c}{N_c-\no} m_Q\mph\,,\quad\langle\Qt\rangle_{\rm br1}\sim\Bigl(\frac{m_Q}{\la}\Bigr )^{\frac{\nt-N_c}{N_c-\no}}\Bigl(\frac{\la}{\mph}\Bigr )^{\frac{\no}{N_c-\no}}\,,\label{(8.1)}
\eq
\bbq
\langle\lym^{\rm (br1)}\rangle^3\equiv\langle S\rangle_{\rm br1}=\frac{ \langle\Qo\rangle_{\rm br1} \langle\Qt\rangle_{\rm br1}}{\mph}\sim \Bigl(\frac{m_Q}{\la}\Bigr )^{\frac{\nt-\no}{N_c-\no}}
\Bigl(\frac{\la}{\mph}\Bigr )_{,}^{\frac{\no}{N_c-\no}}\quad \frac{\langle\Qt\rangle_{\rm br2}}{\langle\Qo\rangle_{\rm br2}}\sim\Bigl (\frac{\mo}{\mph}\Bigr )^{\frac{N_c}{N_c-\no}}\ll 1\,.
\eeq

The potentially important masses look as, see Appendix\,:

a) the quark masses,
\bq
\qma\ll\qmb=\langle m_Q-\Phi_2\rangle=\frac{\langle\Qo\rangle}{\mph}\sim m_Q,\quad {\tilde m}^{\rm pole}_{Q,2}=\frac{\qmb}{z_Q^{+}(\la,\qmb)}\sim \la\Bigl(\frac{m_Q}{\la}\Bigr )^{\nd/N_c}\,;\label{(8.2)}
\eq

b) the gluon masses due to possible higgsing of quarks,
\bbq
(\mgo)^2\sim a_{+}(\mgo)\, z_Q^{+}(\la,\mgo)\,\langle\Qo\rangle,\quad a_{+}(\mgo)=\Bigl (\frac{\la}{\mgo}
\Bigr )^{\nu_{+}},\,\, z_Q^{+}(\la,\mgo)=\Bigl (\frac{\mgo}{\la}\Bigr )^{\gamma^{+}_Q},
\eeq
\bq
\mgo\sim\frac{\langle\Qo\rangle}{\la}\sim\frac{m_Q\mph}{\la},\quad\, \frac{{\tilde m}^{\rm pole}_{Q,2}}{\mgo}\sim\frac{\mo}{\mph}\ll 1,\quad\quad\gamma^{+}_Q=\frac{2N_c-N_F}{N_F-N_c},\quad \nu^{+}=\frac{3N_c-2N_F}{N_F-N_c}\,.\label{(8.3)}
\eq
Therefore, the quarks $Q^1, {\ov Q}_1$ are higgsed and {\bf the overall phase is $\mathbf{Higgs_1-HQ_2}$} in this case. Besides,
\bq
\frac{\mgo}{\mos}\sim\Bigl (\frac{\mph}{\mu_{\Phi,1}}\Bigr )^{\frac{2(2N_c-N_F)}{5N_c-3N_F}}>1\,\,\,
{\rm only \, at}\,\,\, \mph>\mu_{\Phi,1}=\la\Bigl (\frac{\la}{m_Q}\Bigr )^{\frac{5N_c-3N_F}{2(2N_c-N_F)}},\label{(8.4)}
\eq
\bbq
\mos=\la\Bigl (\frac{\la}{\mph}\Bigr )^{\frac{1}{2\gamma_Q^{+}-1}}=\la\Bigl (\frac{\la}{\mph}\Bigl )^{\frac{\nd}{5N_c-3N_F}}\,,\quad\frac{\mo}{\mu_{\Phi,1}}
\sim\Bigl (\frac{m_Q}{\la}\Bigr )^{\frac{\nd(3N_c-2N_F)}{2N_c(2N_c-N_F)}}\ll 1,\quad\mo=\la\Bigl (\frac{\la}{m_Q}\Bigr )^{\frac{2N_c-N_F}{N_c}}\,,
\eeq
this shows that the fions $\Phi^1_1,\Phi_1^2$ and $\Phi_2^1$ become too heavy and dynamically irrelevant at all scales $\mu<\mu_1^{\rm pole}(\Phi)\sim \mph\gg\la$ when $\mph>\mu_{\Phi,1}$ only (the situation with $\Phi^2_2$ is different, see below), while at $\mo<\mph<\mu_{\Phi,1}$ all $N_F^2$ fions become relevant at the scale $\mu<\mu^{\rm pole}_2(\Phi)=\mos=\la(\la/\mph)^{\nd/(5N_c-3N_F)}$ and there appears additional second generation of all $N_F^2\,\,\Phi$-particles.\\

{\bf A)} \,\, The region $\mo\ll\mph\ll\mu_{\Phi,1}=\la (\la/m_Q)^{\frac{5N_c-3N_F}{2(2N_c-N_F)}}$.

Because quarks ${\ov Q}_1, Q^1$ are higgsed at $\mu\sim\mgo$, see \eqref{(8.3)}, the lower energy theory at $\mu<\mgo$ has the unbroken gauge group $SU(N_c-\no)$,\, $n_2$ flavors of quarks ${\ov Q}^{\,\prime}_2, Q^{\,\prime,\,2}$ with unbroken colors, $N_F^2$ fions $\Phi^j_i$, $\,\no^2$ pions $\Pi^1_1$ (these originated from higgsing of ${\ov Q}_1, Q^1$\,), and finally $2\no\nt$ hybrid pions $\Pi^1_2,\, \Pi^2_1$ (these in essence are quarks ${\ov Q}_2, Q^2$ with broken colors).

In case $1\leq \no<{\rm Min}(N_F/2,\,3N_c-2N_F)$, i.e. either any $1\leq \no<N_F/2$ at $N_c<N_F<6N_c/5$, or $\no<(3N_c-2N_F)$ at $6N_c/5<N_F<3N_c/2$, the lower energy theory at $\mu<\mgo$ remains in the strong coupling region with: $N_c^\prime=(N_c-\no),\,\,N_F^\prime=\nt,\,\, 1< N_F^\prime/N_c^\prime<3/2\,,\,\, {\rm b}^\prime_o=(3N_c^\prime-N_F^\prime)=(3N_c-N_F-2\no)>0$.

The anomalous dimensions and the gauge coupling look in this lower energy theory as (see Appendix in \cite{ch6} for the values of anomalous dimensions in the strong coupling region)
\bq
\gamma_Q^{-}=\frac{2N_c^\prime-N_F^\prime}{N_F^\prime-N_c^\prime}=\frac{2N_c-N_F-\no}{(N_F-N_c)=\nd}=
\gamma_Q^{+}-\frac{\no}{\nd}\,>\,1\,,\quad \gamma_{\Phi}^{\pm}=-2\gamma_Q^{\pm}\,,\label{(8.5)}
\eq
\bbq
a_{-}(\mu<\mgo)=\Bigl [a_{+}(\mu=\mgo)=\Bigl (\frac{\la}{\mgo}\Bigr )^{\nu_{+}}\Bigr ]\Bigl (\frac{\mgo}{\mu}\Bigr )^{\nu_{-}},\quad \nu_{+}=\frac{3N_c-2N_F}{\nd},\quad \nu_{-}=\nu_{+}-\frac{\no}{\nd}>0\,.
\eeq
The pole mass of ${\ov Q}^{\,\prime}_2, Q^{\,\prime,\,2}$ quarks and $\mgt$ due to possible higgsing of these quarks look then as, see \eqref{(8.1)},\eqref{(8.3)},
\bbq
m_{Q,2}^{\rm pole}=\frac{\qmb}{z^{+}_Q(\la,\mgo)\,z^{-}_Q(\mgo,m_{Q,2}^{\rm pole})}\,,\,\, z^{+}_Q(\la,\mgo)=\Bigl (\frac{\mgo}{\la}\Bigr )^{\gamma^{+}_Q},\,\, z^{-}_Q(\mgo,m_{Q,2}^{\rm pole})=\Bigl (\frac{m_{Q,2}^{\rm pole}}{\mgo}\Bigr )^{\gamma^{-}_Q}\,,
\eeq
\bq
m_{Q,2}^{\rm pole}\sim\Bigl (\frac{\la}{\mph}\Bigr )^{\frac{\no}{N_c-\no}}\,\Bigl (\frac{m_Q}{\la}\Bigr )^{\frac{\nd-\no}{N_c-\no}}\sim\frac{\langle\Qt\rangle}{\la}\,,\quad\quad \frac{m_{Q,2}^{\rm pole}}
{\mgo}\sim\Bigl (\frac{\mo}{\mph}\Bigr )^{\frac{N_c}{N_c-\no}}\ll 1\,,\label{(8.6)}
\eq
\bbq
(\mgt)^2\sim \rho_{+}\rho_{-}\langle\Qt\rangle,\,\, \rho_{+} =a_{+}(\mgo)\, z_Q^{+}(\la,\mgo)
\sim\frac{\mgo}{\la},\,\, \rho_{-} \sim\Bigl (\frac{\mgo}{\mgt}\Bigr )^{\nu^-} z_Q^{-}(\mgo,\mgt)=\frac{\mgt}{\mgo}\,,
\eeq
\bbq
\rho_{+}\rho_{-}\sim\frac{\mgt}{\la}\,,\quad \mgt\sim m_{Q,2}^{\rm pole}\sim \frac{\langle\Qt\rangle}
{\la}\sim \Bigl(\frac{m_Q}{\la}\Bigr )^{\frac{\nt-N_c}{N_c-\no}}\Bigl(\frac{\la}{\mph}\Bigr )^{\frac{\no}{N_c-\no}}\,.
\eeq
Because $\nt-(N_c^\prime=N_c-\no)=N_F-N_c>0$, the rank restriction shows that the qurks  ${\ov Q}^{\,\prime}_2, Q^{\,\prime,\,2}$ are not higgsed because otherwise the global $U(\nt)$ flavor symmetry will be {\it additionally} broken spontaneously (and this will be wrong).

Therefore, after integrating out remained active quarks ${\ov Q}^{\,\prime}_2, Q^{\,\prime,\,2}$ with unbroken $N_c-\no$ colors as heavy ones at $\mu<m_{Q,2}^{\rm pole}$, the scale factor of the lower energy $SU(N_c)$ SYM in the strong coupling regime is determined from the matching, see \eqref{(8.3)},\eqref{(8.5)},\eqref{(8.6)},
\bq
a_{-}(\mu=m_{Q,2}^{\rm pole})=\Bigl (\frac{\la}{\mgo}\Bigr )^{\nu_{+}}\,\Bigl (\frac{\mgo}{m_{Q,2}^{\rm pole}}\Bigr )^{\nu_{-}}=a_{\rm YM}^{(\rm br1)}(\mu=m_{Q,2}^{\rm pole})=\Bigl (\frac{m_{Q,2}^{\rm pole}}{\lambda_{YM}}\Bigr )^3\gg 1\,,\label{(8.7)}
\eq
and is
\bq
\lambda_{YM}^3=\Bigl (\langle\lym^{(\rm br1)}\rangle\Bigr )^3=\la^3\Bigl (\frac{m_Q}{\la}\Bigr )^{\frac{\nt-\no}{N_c-\no}}\Bigl (\frac{\la}{\mph}\Bigr )^{\frac{\no}{N_c-\no}}=\langle S\rangle_{\rm br1}=\frac{\langle\Qo\rangle_{\rm br1}\langle\Qt\rangle_{\rm br1}}{\mph}\,,\label{(8.8)}
\eq
as it should be, see \eqref{(8.1)}, while
\bq
\frac{\langle\lym^{(\rm br1)}\rangle}{m_{Q,2}^{\rm pole}}=\Bigl (\frac{m_Q}{\la}\Bigr )^{\frac{3N_c-2N_F+\no}{3(N_c-\no)}}\Bigl (\frac{\mph}{\la}\Bigr )^
{\frac{2\no}{3(N_c-\no)}}=\Bigl (\frac{\mph}{{\mu}_{\Phi,3}}\Bigr )^{\frac{2\no}{3(N_c-\no)}}\ll 1 \label{(8.9)}
\eq
\bbq
{\rm at}\,\quad \mph\ll {\rm Min}\,\Bigl\{\frac{\la^2}{m_Q},\,{\mu}_{\Phi,3}\Bigr\},\quad
{\mu}_{\Phi,3}=\la(\frac{\la}{m_Q}\Bigr )^{\frac{3N_c-2N_F+\no}{2\no}}\,,
\eeq
(at $\mph>\la^2/m_Q$ the quarks are higgsed in the weak coupling regime, see \eqref{(8.1)}).

Integrating out finally all $SU(N_c-\no)$ gluons at $\mu<\langle\lym^{(\rm br1)}\rangle\ll m_{Q,2}^{\rm pole}$ via the VY-procedure \cite{VY}, the lower energy Lagrangian of $N_F^2$ 3-rd generation fions $\Phi_i^j$,\, $\no^2$ pions $\Pi^i_j$ and $2\no\nt$ hybrid pions $\Pi^1_2,\, \Pi^2_1$ (in essence, these are ${\ov Q}_2, Q^2$ quarks with broken colors) look as, see section 6.1 for the form of $K_{\Pi}$,
\bq
K_{\rm tot}=K_{\Phi}+K_{\Pi}+K^{\rm hybr}_{\Pi}\,,\label{(8.10)}
\eq
\bbq
K_{\Phi}=z_{\Phi}^{+}(\la,\mgo)\,{\rm Tr}\,\Bigl [(\Phi_1^1)^\dagger\Phi_1^1+\Bigl (\Phi_1^2)^\dagger\Phi_1^2+(\Phi_2^1)^\dagger\Phi_2^1\Bigr )+z_{\Phi}
^{-}(\mgo,m_{Q,2}^{\rm pole})\,(\Phi_2^2)^\dagger\Phi_2^2\Bigr ]\,,
\eeq
\bbq
K_{\Pi}\simeq K^{\rm (loop)}_{\Pi} =\,{\rm Tr}\,\frac{(\Pi_1^1)^\dagger\Pi_1^1}{\la^2}\,,\quad
K^{\rm hybr}_{\Pi}=z_{Q}^{+}(\la,\mgo)\,{\rm Tr}\,\Bigl (\,\,\frac{(\Pi_1^2)^\dagger\Pi_1^2+(\Pi_2^1)^\dagger
\Pi_2^1}{|\langle\Pi_1\rangle|=|\langle\Qo\rangle|}\,\,\Bigr )\,,
\eeq
\bq
{\cal W}_{\rm tot}={\cal W}_{\Phi}+{\cal W}_{\Phi\Pi}+{\cal W}_{\rm non-pert}\,,\label{(8.11)}
\eq
\bbq
{\cal W}_{\Phi}=\frac{\mph}{2}\Bigl [\,{\rm Tr}\, (\Phi^2)-\frac{1}{\nd}\,\Bigl ({\rm Tr}\,\Phi\Bigr )
^2\Bigr ]\,,\quad {\cal W}_{\rm non-pert}=(\nd-\no)\Bigl (\frac{\la^{\rm \bo}\det m_{Q,2}^{\rm tot}}{\det\Pi_1^1}\Bigr )^{\frac{1}{N_c-\no}}\,,
\eeq
\bbq
{\cal W}_{\Phi\Pi}=m_Q\,{\rm Tr}\,\Bigl [ \Pi_1^1+\Pi^2_1\frac{1}{\Pi_1^1}\Pi^1_2\Bigr ]-
\,{\rm Tr}\,\Bigl [\, \Pi_1^1\Phi_1^1+\Pi_1^2\Phi_2^1
+\Pi_2^1\Phi_1^2+\Phi_2^2\,(\Pi^2_1\frac{1}{\Pi_1^1}\Pi^1_2)\,\Bigr ]\,.
\eeq

We obtain from \eqref{(8.10)},\eqref{(8.11)} for the particle masses. -

The masses of the 3-rd generation fions $\mu^{\rm pole}_{3}(\Phi_1^1)\sim\mu^{\rm pole}_{3}(\Phi_1^2)\sim\mu^{\rm pole}_{3}(\Phi_2^1)$ look as
\bq
\mu^{\rm pole}_{3}(\Phi_1^1)=\frac{\mph}{z_{\Phi}^{+}(\la,\mgo)}\sim\la\Bigl (\frac{m_Q}
{\la}\Bigr )^{2(2N_c-N_F)/\nd}\Bigl (\frac{\mph}{\la}\Bigr )^{(3N_c-N_F)/\nd}\,,\label{(8.12)}
\eq
\bbq
\frac{\mu^{\rm pole}_{3}(\Phi_1^1)}{\mgo}\sim\Bigl (\frac{\mph}{\mu_{\Phi,1}}\Bigr )^{2(2N_c-N_F)/\nd}
\ll 1\,,
\eeq
while for $\mu^{\rm pole}_{3}(\Phi_2^2)$,
\bq
\mu^{\rm pole}_{3}(\Phi_2^2)\sim\frac{\mph}{z_{\Phi}^{+}(\la,\mgo)z_{\Phi}^{-}(\mgo,m_{Q,2}^{\rm pole})}\sim
\la\Bigl (\frac{m_Q}{\la}\Bigr )^{\frac{2(2N_c-N_F)}{N_c-\no}}\,
\Bigl (\frac{\mph}{\la}\Bigr )^{\frac{N_c+\no}{N_c-\no}}\,,\label{(8.13)}
\eq
\bbq
\frac{\mu^{\rm pole}_{3}(\Phi_2^2)}{\mu^{\rm pole}_{3}(\Phi_1^1)}=\frac{1}{z_{\Phi}^{-}(
\mgo,m_{Q,2}^{\rm pole})}\sim\Bigl (\frac{m_{Q,2}^{\rm pole})}{\mgo}\Bigr )^{2\gamma^{-}_Q}\sim\Bigl (\frac{\mo}{\mph}\Bigr )^{\frac{2N_c}{N_c-\no}(\gamma^{-}_Q\,>\,1)}\ll 1\,,
\eeq
\bbq
\frac{\mu^{\rm pole}_{3}(\Phi_2^2)}{m_{Q,2}^{\rm pole}}\sim\Bigl (\frac{\mph}{\mu_{\Phi,2}} \Bigr )^{\frac{N_c+2\no}{N_c-\no}}\ll 1\quad {\rm at}\quad \mph\ll\mu_{\Phi,2},\quad\mu_{\Phi,1}\ll
\mu_{\Phi,2}=\la\Bigl (\frac{\la}{m_Q}\Bigr )^{\frac{5N_c-3N_F+\no}{N_c+2\no}}
\eeq
(the main contribution to $\mu^{\rm pole}_{3}(\Phi_2^2)$ in \eqref{(8.13)} originates from the term $\sim \mph{\rm Tr}\,(\Phi_2^2)^2$ in \eqref{(8.11)}, while the contribution from ${\cal W}_{\rm non-pert}$ is much smaller\,).

The smallest nonzero masses have $\no^2$ pions $\Pi^1_1$,
\bq
\mu^{\rm pole}(\Pi_1^1)\sim\frac{\la^2}{\mph}\,,\quad\quad \frac{\mu^{\rm pole}(\Pi_1^1)}{\mu^{\rm pole}_{3}(\Phi_2^2)}\sim\Bigl (\frac{\mo}{\mph}\Bigr )^{\frac{2N_c}{N_c-\no}}\ll 1 \label{(8.14)}
\eq
(the main contribution to $\mu^{\rm pole}(\Pi_1^1)$ in \eqref{(8.14)} originates from the mixing term $\sim  {\rm Tr}\,(\Pi_1^1\Phi_1^1)$ in \eqref{(8.11)}, after integrating out heavier $\Phi^1_1$, while the contribution to $\mu^{\rm pole}(\Pi_1^1)$  from ${\cal W}_{\rm non-pert}$ is much smaller\,).

Finally, $2\no\nt$ multiplets $\Pi_1^2$ and $\Pi_2^1$ are the Nambu-Goldstone particles and are massless.\\

{\bf B)}\,\, The region $\mu_{\Phi,1}=\la (\la/m_Q)^{\frac{5N_c-3N_F}{2(2N_c-N_F)}}\ll\mph\ll\la^2/m_Q$\,.

Because $\mgo>\mos=\la(\la/\mph)^{\nd/(5N_c-3N_F)}$ in this case, the running mass $\mu_{\Phi}(\mu)$ of all $N_F^2$ fions $\Phi^j_i$ remains too large at the scale $\mu=\mgo,\,\, \mu_{\Phi}(\mu=\mgo)>\mgo$, so that they are still irrelevant at this scale. Because all heavy higgsed quarks ${\ov Q}_1, Q^1$ decouple at $\mu<\mgo$, the RG evolution of fions $\Phi_1^1, \Phi_1^2$ and $\Phi_2^1$ is frozen at scales $\mu<\mgo$, so that they remain irrelevant at all scales $\mu<\mu_1^{\rm pole)}(\Phi)\sim\mph\gg\la$, there are no additional generations of these particles.

But as for $\Phi_2^2$\,, the situation is different because the quarks ${\ov Q}^{\,\prime}_2, Q^{\,\prime,\,2}$ with unbroken colors are still active at $\mu<\mgo$ and the running mass of fions $\Phi_2^2$ continues to decrease with decreasing scale (until ${\ov Q}^{\,\prime}_2, Q^{\,\prime,\,2}$ decouple at $\mu<m_{Q,2}^{\rm pole}$). As a result, there still will be two additional generations of $n_2^2\,\, \Phi_2^2$-particles in the range $\mu_{\Phi,1}<\mph<\mu_{\Phi,2}<\la^2/m_Q$, see \eqref{(8.13)},
\bq
\mu^{\rm pole}_{2}(\Phi_2^2)={\hat\mu}_o^{\,\rm str}=\la\Bigl (\frac{\la}{{\hat\mu}_{\Phi}}\Bigr )^{\frac{1}{2\gamma^{-}_Q-1}}=\la\Bigl (\frac{\la}{m_Q}\Bigr )^{\frac{2n_1}{5N_c-3N_F-2\no}}\Bigl (\frac{\la}{\mph}\Bigr )^{\frac{\nd+2\no}{5N_c-3N_F-2\no}}\,,\label{(8.15)}
\eq
\bbq
{\hat\mu}_{\Phi}=\mph\Bigl (\frac{\mgo}{\la}\Bigr )^{2(\gamma_Q^{+}-\gamma_Q^{-})}=
\la\Bigl(\frac{m_Q}{\la}\Bigr )^{\frac{2\no}{\nd}}\Bigl (\frac{\mph}{\la}\Bigr )^{\frac{\nd+2\no}{\nd}},
\quad \frac{\mu^{\rm pole}_{2}(\Phi_2^2)}{\mgo}=\Bigl(\frac{\mu_{\Phi,1}}{\mph}\Bigr )^{\frac{2(2N_c-N_F)}{5N_c-3N_F-2\no}\,>\,0}\ll 1\,.
\eeq
\bq
\frac{m_{Q,2}^{\rm pole}}{\mu^{\rm pole}_{2}(\Phi_2^2)}\sim\Bigl (\frac{\mph}{\mu_{\Phi,2}}\Bigr )^{\frac{\nd(N_c+2\no)}{(N_c-\no)(5N_c-3N_F-2\no)}>\,0}\,,\quad
\quad\frac{\mu_{\Phi,1}}{\mu_{\Phi,2}}\sim\Bigl (\frac{m_Q}{\la}\Bigr )^{\frac{(3N_c-2N_F)(5N_c-3N_F-2\no)}{2(2N_c-N_F)(N_c+2\no)}\,>\,0}\ll 1\,.\label{(8.16)}
\eq

As it is seen from \eqref{(8.16)}, all $\nt^2$ fions $\Phi_2^2$ remain dynamically irrelevant only at $\mph>\mu_{\Phi,2}\gg\mu_{\Phi,1}\gg\mo$, because the quarks ${\ov Q}^{\,\prime}_2, Q^{\,\prime,\,2}$  with unbroken colors decouple in this case before there appears the second generation of $\Phi_2^2$\,, while in the range $\mu_{\Phi,1}\ll\mph\ll\mu_{\Phi,2}$ there appear two additional generations of $\nt^2\,\, \Phi_2^2$-particles, and fions $\Phi_2^2$ become dynamically relevant in the interval of scales $\mu^{\rm pole}_{3}(\Phi_2^2)<\mu<\mu^{\rm pole}_{2}(\Phi_2^2)$, see \eqref{(8.13)},\eqref{(8.15)}.

But to keep $\mu_{\Phi,2}<\la^2/m_Q$ (the RG flow becomes different at $\mph>\la^2/m_Q$ because $\mgo\sim (m_Q\mph)/\la$ becomes larger than $\la$ and the quarks ${\ov Q}_1, Q^1$ will be higgsed in the weak coupling regime) it should be:  $8N_c/7<N_F<3N_c/2$. Therefore, at $N_c<N_F<8N_c/7$ the $\nt^2$ fions $\Phi_2^2$
are relevant in the range of scales $\mu^{\rm pole}_{3}(\Phi_2^2)<\mu<\mu^{\rm pole}_{2}(\Phi_2^2)$ in the whole interval $\mo<\mph<\la^2/m_Q$ and there are two additional generations, $\mu^{\rm pole}_2(\Phi_2^2)$ and $\mu^{\rm pole}_3(\Phi_2^2)$, see \eqref{(8.4)},\eqref{(8.13)},\eqref{(8.15)}. But at $8N_c/7<N_F<3N_c/2$ these two additional generations of particles $\Phi_2^2$ exist only in the range $\mu_{\Phi,1}<\mph<\mu_{\Phi,2}$ and disappear at $\mu_{\Phi,2}<\mph<\la^2/m_Q$, so that there are no additional generations of any $\Phi$-particles at all at $\mph>\mu_{\Phi,2}$\,.\\

All formulas in special vacua with $\no=\nd,\, \nt=N_c$ can be obtained simply substituting $\no=\nd$ in all expressions in this section 8.1.

\subsection{\quad  br1 and special vacua with $3N_c-2N_F < n_1< N_F/2$}

Consider now the case $\no > 3N_c-2N_F$, this requires $6N_c/5< N_F < 3N_c/2$ since $\no < N_F/2$ (recall also that we ignore for simplicity all RG evolution effects if they are logarithmic only). In this case $3/2 < N_F^\prime/N_c^\prime < 3\,,\,\,\, N_F^\prime=N_F-\no=\nt\,,\,\,\,N_c^\prime=
N_c-\no$. This means that, after the quarks ${\ov Q}_1, Q^1$ are higgsed at $\mu\sim\mgo\sim m_Q\mph/\la\ll\la$ in the strong coupling regime $a_{+}(\mu=\mgo)\gg 1$, the theory remains with the same large coupling $a_{-}(\mu=\mgo)=a_{+}(\mu=\mgo)$, but the numbers of colors $N_c^\prime=N_c-\no$ and flavors $N_F^\prime=N_F-\no=\nt$ in the lower energy theory are now such that the coupling $a_{-}(\mu)$ begins {\it to decrease}  with diminishing scale because the $\beta$-function becomes {\it positive} at $\mu<\mgo$ (until the gauge coupling is large, $a_{-}(\mu)\gg 1$), while the quark anomalous dimension looks now as $\gamma^{-}_Q(N^\prime_c,N^\prime_F,a(\mu)\gg 1)=(2N^\prime_c-N^\prime_F)/(N^\prime_F-N^\prime_c)$,
\bq
\frac{d a_{-}(\mu)}{d\ln\mu}=\frac{a^2_{-}(\mu)[\, 3-\frac{N_F^\prime}{N_c^\prime}(1+\gamma^{-}_Q) ]}
{a_{-}(\mu)-1}\simeq a_{-}(\mu)\Bigl [\frac{\,\no-(3N_c-2N_F)}{\nd}\Bigr ]>0, \label{(8.17)}
\eq
\bbq
0\,<\,\gamma^{-}_Q=\frac{2N_c-N_F-\no}{\nd}\,<\, 1,\,\, \gamma^{-}_{\Phi}=-2\gamma^{-}_Q \,\,\, {\rm at}
\,\,\, (3N_c-2N_F)<\no<(2N_c-N_F)\,,\,\, \frac{6N_c}{5}< N_F < \frac{4N_c}{3}\,,
\eeq
\bbq
-\,\frac{1}{2}<\,\gamma^{-}_Q=\frac{2N_c-N_F-\no}{\nd}\,<\,0,\,\, \gamma^{-}_{\Phi}=0 \,\,\, {\rm at}\,\,\, (2N_c-N_F)<\no<\frac{N_F}{2}\,,\quad \frac{4N_c}{3}< N_F < \frac{3N_c}{2}\,.
\eeq

The scale factor $\Lambda^\prime_Q$ of the gauge coupling $a_{-}(\mu)$ is determined from the matching
\bbq
a_{+}(\mu=\mgo)=\Bigl (\frac{\la}{\mgo}\Bigr )^{\frac{3N_c-2N_F}{N_F-N_c}}=a_{-}(\mu=\mgo)=\Bigl (\frac{\mgo}{\Lambda^\prime_Q}\Bigr )^{\frac{\no-(3N_c-2N_F)}{N_F-N_c}}\gg 1\,,
\eeq
\bq
\Lambda^\prime_Q=\la\Bigl (\frac{m_Q\mph}{\la^2}\Bigr )^{\frac{\no}{\no-(3N_c-2N_F)}\,>\, 1}\ll\mgo\sim \frac{m_Q\mph}{\la}\ll\la\,.\label{(8.18)}
\eq
Therefore (if nothing prevents, see below), the theory will finally enter smoothly at the scale $\mu<\Lambda^\prime_Q$ into the conformal regime with the frozen gauge coupling $a_{*}\sim 1$. But, if $m_{Q,2}^{\rm pole}\gg\Lambda^\prime_Q$, see \eqref{(8.9)}, the quarks ${\ov Q}^{\,\prime}_2, Q^{\,\prime,\,2}$  with unbroken colors will decouple still in the strong coupling regime $a_{-}(\mu=m_{Q,2}^{\rm pole})\gg 1$ and the lower energy gauge theory will be ${\cal N}=1\,\, SU(N_c-\no)$ SYM in the strong coupling regime, with the scale factor of its gauge coupling determined as previously from the matching
\bbq
a_{-}(\mu=m_{Q,2}^{\rm pole})=\Bigl (\frac{m_{Q,2}^{\rm pole}}{\Lambda^\prime_Q}\Bigr )^{\frac{\no-(3N_c-2N_F)}{N_F-N_c}\,>\,0}=a^{(\rm br1)}_{YM}(\mu=m_{Q,2}^{\rm pole})=
\Bigl (\frac{m_{Q,2}^{\rm pole}}{\lambda_{YM}}\Bigr )^{3}\quad\ra
\eeq
\bq
\ra\quad\lambda^3_{YM}=\Bigl (\langle\lym^{(\rm br1)}\rangle\Bigr )^3=\la^3\Bigl (\frac{m_Q}{\la}\Bigr )^{\frac{\nt-\no}{N_c-\no}}\Bigl (\frac{\la}{\mph}\Bigr )^{\frac{\no}{N_c-\no}}=\langle S\rangle_{\rm br1}. \label{(8.19)}
\eq
as it should be, see \eqref{(8.1)},\eqref{(8.8)}.

Because our main purpose in this article is to calculate the mass spectra in the direct theory in cases with quarks in the strong coupling regime $a(\mu)\gg 1$, we consider below this case $m_{Q,2}^{\rm pole}\gg\Lambda^\prime_Q$ only. This requires then, see \eqref{(8.9)},
\bq
\frac{\Lambda^\prime_Q}{m_{Q,2}^{\rm pole}}\ll 1\quad\ra\quad \mph\ll {\mu}_{\Phi,3}=\la\Bigl (\frac{\la}{m_Q}\Bigr )^{\frac{3N_c-2N_F+\no}{2\no}\,<\,1}<\frac{\la^2}{m_Q}\,,\quad \no>3N_c-2N_F\,.\label{(8.20)}
\eq

We distinguish then three regions. -\\

{\bf I)} $6N_c/5\,<\,N_F\,<\,5N_c/4,\,\,(3N_c-2N_F)\,<\,\no\,<\,N_F/2$.  Here the hierarchies look as: $\,\mo\ll\mu_{\Phi,1}\ll\mu_{\Phi,2}\ll\mu_{\Phi,3}\ll\la^2/m_Q$, see \eqref{(8.4)},\eqref{(8.13)},\eqref{(8.20)}, while $1/2\,<\gamma_Q^{-}\,<1$.\\
{\bf a)}\, The region $\mo\ll\mph\ll\mu_{\Phi,1}$. There are two complete additional generations of all $N^2_F$ $\Phi$-particles, $\Phi_{1}^1, \Phi_{1}^2, \Phi_{2}^1$ are relevant in the range of scales $\mu_3^{\rm pole}(\Phi^1_1)<\mu<\mu_2^{\rm pole}(\Phi^1_1)$, while $\Phi_{2}^2$ is relevant at $\mu_3^{\rm pole}(\Phi^2_2)<\mu<\mu_2^{\rm pole}(\Phi^2_2),\,\,\mu_2^{\rm pole}(\Phi^2_2)=\mu_2^{\rm pole}(\Phi^1_1)=\mos=\la(\la/m_Q)^{\nd/(5N_c-3N_F)}$, see \eqref{(8.4)},\eqref{(8.12)},\eqref{(8.13)}.\\
{\bf b)}\, The region $\mu_{\Phi,1}\ll\mph\ll\mu_{\Phi,2}$. $\,\,\,\Phi_{1}^1, \Phi_{1}^2, \Phi_{2}^1$ are irrelevant at all scales $\mu<\mu^{\rm pole}_1(\Phi)\sim\mph$. As for $\Phi_2^2$, there still will be two additional generations of $\Phi_2^2$-particles with masses $m_{Q,2}^{\rm pole}<\mu^{\rm pole}_{2}
(\Phi_2^2)<\mgo$, see \eqref{(8.15)},\eqref{(8.16)}, and $\mu^{\rm pole}_{3}(\Phi_2^2)<m_{Q,2}^{\rm pole}$, see \eqref{(8.13)}.\\
{\bf c)}\, The region $\mu_{\Phi,2}\ll\mph\ll\mu_{\Phi,3}$. In this case all $N^2_F$ $\Phi$-particles are irrelevant at all scales $\mu<\mu^{\rm pole}_1(\Phi)\sim\mph$.\\

{\bf II)} $5N_c/4\,<\,N_F\,<\,4N_c/3,\,\, (5N_c-3N_F)/2\,<\,\no\,<\,N_F/2$. Here the hierarchies look as: $\mo\ll\mu_{\Phi,2}\ll\mu_{\Phi,1}\ll\mu_{\Phi,3}\ll\la^2/m_Q$, while $0\,<\gamma_Q^{-}\,<\,1/2$.\\
{\bf a)}\, The region $\mo\ll\mph\ll\mu_{\Phi,2}$. There will be 2-nd and 3-rd generations of all $N_F^2\,\, \Phi^j_i$-particles with $\mu^{\rm pole}_2(\Phi)=\mos=\la(\la/m_Q)^{\nd/(5N_c-3N_F)}$ and $\mu^{\rm pole}_3(\Phi_1^1)$ in \eqref{(8.12)}, $\mu^{\rm pole}_3(\Phi_2^2)<m_{Q,2}^{\rm pole}$ in \eqref{(8.13)}.\\
{\bf b)}\, The region $\mu_{\Phi,2}\ll\mph\ll\mu_{\Phi,1}$. The difference with "{a}"\, is that now there is no third generation of fions $\mu_3(\Phi_2^2)$.\\
{\bf c)}\, The region $\mu_{\Phi,1}\ll\mph\ll\mu_{\Phi,3}$. All fions are too heavy and irrelevant in this case at all scales $\mu<\mu_1^{\rm pole}\sim\mph\gg\la$. \\

{\bf III)} $4N_c/3\,<\,N_F\,<\,3N_c/2,\,\, (2N_c-N_F)\,<\,\no\,<\,N_F/2$. Because $(-1/2)\,<\gamma_Q^{-}\,
<\,0$ and $z_{\Phi}^{-}=1$ in \eqref{(8.10)} now, then $\mu^{\rm pole}_3(\Phi_2^2)\sim\mu^{\rm pole}_3(\Phi_1^1)$ \eqref{(8.12)} instead of \eqref{(8.13)}, while the hierarchies look now as: $\mo\ll\mu_{\Phi,3}\ll\mu_{\Phi,1}\ll\la^2/m_Q$. Because we consider only the region $\mph<\mu_{\Phi,3}$ and $\mu_{\Phi,3}<\mu_{\Phi,1}$ now, there will be 2-nd and 3-rd generations of all $N_F^2\,\, \Phi^j_i$-particles with $\mu^{\rm pole}_2(\Phi^j_i)=\mos=\la(\la/\mph)^{\nd/(5N_c-3N_F)}$ \eqref{(8.4)}, and $\mu^{\rm pole}_3(\Phi^j_i)$ in \eqref{(8.12)}.\\

All formulas in special vacua with $\no=\nd,\, \nt=N_c$ can be obtained simply substituting $\no=\nd$ in all expressions in this section 8.2.

\subsection{\quad  br2-vacua}

In these br2 vacua with $N_F/2<\nt<N_c$, all expressions for condensates and corresponding masses in  \eqref{(8.1)}-\eqref{(8.4)} are obtained replacing $\no\leftrightarrow\nt$. Because we are interested in this article in that the direct theory stays in the strong coupling regime $a(\mu<\la)\gtrsim 1$, all other formulae for br2-vacua can also be obtained from those for br1-vacua by $\no\leftrightarrow\nt$, under corresponding restrictions.

\section{Broken flavor symmetry.\, Dual theory. }

\subsection{\quad  br1 and special vacua with $1\leq n_1<{\rm Min}\,(N_F/2,\,3N_c-2N_F)$}

This dual theory is mostly in the IR free logarithmic regime in this case and all logarithmic effects of the RG evolution will be ignored below for simplicity, as before. All $N_F^2$ fions $\Phi^j_i$ have very large masses $\sim\mph\gg\la$ and can be integrated out from the beginning. The potentially most important masses of dual quarks and gluons look here as follows, see Appendix,
\bq
\omp\sim\frac{\langle M_1\rangle_{\rm br1}=\langle\Qo\rangle_{\rm br1}}{\la}\sim\frac{m_Q\mph}{\la}\gg\tmp\sim\frac{\langle M_2\rangle_{\rm br1}}{\la}
\sim\la\Bigl (\frac{m_Q}{\la}\Bigr )^{\frac{\nd-n_1}{N_c-n_1}}\Bigl (\frac{\la}{\mph}\Bigr )^{\frac{n_1}{N_c-n_1}}\,,\label{(9.1)}
\eq
\bbq
\langle\lym^{(\rm br1)}\rangle^3\equiv\langle S\rangle_{\rm br1}=\frac{\langle M_1\rangle\langle M_2\rangle}{\mph}\sim\la^3\Bigl (\frac{m_Q}{\la}\Bigr )^{\frac{\nt-n_1}{N_c-n_1}}\Bigl (\frac{\la}{\mph}\Bigr )^{\frac{n_1}{N_c-n_1}}\,,
\eeq
while, see \eqref{(9.1)},
\bq
({\ov\mu}^{\,\rm pole}_{gl,2})^2\sim\langle\qt\rangle_{\rm br1}=\frac{\langle S\rangle_{\rm br1}\la}{\langle M_2\rangle_{\rm br1}}=\frac{\langle M_1\rangle_{\rm br1}\la}{\mph}\sim m_Q\la\gg (\mgo)^2\,,\label{(9.2)}
\eq
\bq
\Bigl (\frac{{\ov\mu}^{\,\rm pole}_{gl,2}}{\tmp}\Bigr )^2\sim\Bigl (\frac{m_Q}{\la}\Bigr )^{\frac{3N_c-2N_F+n_1}{N_c-n_1}}\Bigl (\frac{\mph}{\la}\Bigr )^{\frac{2n_1}{N_c-n_1}}\ll 1\,,\label{(9.3)}
\eq
\bbq
\Bigl (\frac{{\ov\mu}^{\,\rm pole}_{gl,2}}{\omp}\Bigr )^2\sim\frac{\la^3}{m_Q\mph^2}
<\frac{\la^3}{m_Q\mo^2}\sim\Bigl (\frac{m_Q}{\la}\Bigr )^{\frac{3N_c-2N_F}{N_c}}\ll 1\,,
\eeq
\bbq
\Bigl (\frac{\langle\lym^{(\rm br1)\rangle}}{\tmp}\Bigr )^3\sim\Bigl (\frac{{\ov\mu}^{\,\rm pole}_{gl,2}}{\tmp}\Bigr )^2\sim\Bigl (\frac{m_Q}{\la}\Bigr )^{\frac{3N_c-2N_F+n_1}{N_c-n_1}}\Bigl (\frac{\mph}{\la}\Bigr )^{\frac{2n_1}{N_c-n_1}}<\Bigl (\frac{m_Q}{\la}\Bigr )^{\frac{3N_c-2N_F-n_1}{N_c-n_1}}\ll 1\,.
\eeq

Therefore, {\bf the overall phase is $\mathbf{Hq_1-Hq_2}$}, all dual quarks are not higgsed but confined, the confinement originates from the unbroken $SU(\nd)\,\,{\cal N}=1$ SYM and so the scale of the confining string tension is $\sqrt\sigma\sim\lym^{(\rm br1)}=\langle S\rangle^{1/3}_{\rm br1}$, see \eqref{(9.1)}.

At scales $\mu<\omp$ all ${\ov q}^1, q_1$ quarks can be integrated out as heavy ones and there remain $SU(\nd)$ gauge theory with $n_2$ flavors of quarks ${\ov q}^2, q_2$ (and with $\bd^{\,\prime}=
2N_F-3N_c+n_1\,<0\,)$ and $N_F^2$ mions $M^i_j$. Integrating out then these quarks as heavy ones at $\mu<\tmp$ and then all ${\cal N}=1\,\, SU(\nd)$ SYM-gluons at $\mu<\lym^{(\rm br1)}$ via the VY-procedure, the lower energy Lagrangian of mions looks as
\bq
K\sim {\rm Tr}\,\frac{M^\dagger M}{\la^2}\,,\quad \quad{\cal W}={\cal W}_M+{\cal W}_{\rm non-pert}\,,\quad
{\cal W}_{\rm non-pert}= -\nd\Bigl (\frac{\det M}{\la^{\rm\bo}}\Bigr )^{1/\nd}\,,\label{(9.4)}
\eq
\bbq
{\cal W}_M=-\frac{1}{2\mph}\Bigl [{\rm Tr}\,(M^2)-\frac{1}{N_c}\Bigl ({\rm Tr}\, M\Bigr )^2\Bigr ].
\eeq
From \eqref{(9.4)}, the masses of mions $M^i_j$ look as follows\,:\\
1)\,\, $n_2^2$ mions $M_2^2$ have masses
\bq
\mu^{\rm pole}(M_2^2)\sim\frac{\langle M_1\rangle_{\rm br1}}{\langle M_2\rangle_{\rm br1}}\,\frac{\la^2}{\mph}\sim\la\Bigl (\frac{m_Q}{\la}\Bigr )^{\frac{2N_c-N_F}{N_c-n_1}}\Bigl (\frac{\mph}{\la}\Bigr )^{\frac{n_1}{N_c-n_1}}\,,\label{(9.5)}
\eq
\bbq
\Bigl (\frac{\mu^{\rm pole}(M_2^2)}{\lym^{(\rm br1)}}\Bigr )^{3/2}\sim\Bigl (\frac{m_Q}{\la}\Bigr )^{\frac{3N_c-2N_F+n_1}{N_c-n_1}}\Bigl (\frac{\mph}{\la}\Bigr )^{\frac{2n_1}{N_c-n_1}}<\Bigl (\frac{m_Q}{\la}\Bigr )^{\frac{3N_c-2N_F-n_1}{N_c-n_1}}\ll 1\,,
\eeq
the main contribution to $\mu^{\rm pole}(M_2^2)$ originates from ${\cal W}_{\rm non-pert}$ in \eqref{(9.4)}.

2)\,\, $n_1^2$ mions $M_1^1$ have masses
\bq
\mu^{\rm pole}(M_1^1)\sim\frac{\la^2}{\mph}\sim\frac{\langle M_2\rangle_{\rm br1}}{\langle M_1\rangle_{\rm br1}}\,\mu^{\rm pole}(M_2^2)\sim\Bigl (\frac{\mo}{\mph}\Bigr )^{\frac{N_c}{N_c-n_1}}\,\mu^{\rm pole}(M_2^2)\ll \mu^{\rm pole}(M_2^2)\,,\label{(9.6)}
\eq
the main contribution to $\mu^{\rm pole}(M_1^1)$ originates from ${\cal W}_{M}$ in \eqref{(9.4)}.

3)\,\, $2n_1n_2$ mions $M_1^2, M_2^1$ are the Nambu-Goldstone particles and are massless
\bq
\mu^{\rm pole}(M_1^2)= \mu^{\rm pole}(M_2^1)=0\,.\label{(9.7)}
\eq

On the whole, the mass spectrum looks in this case as follows. -\\
1) There is a large number of hadrons made from weakly interacting non-relativistic dual quarks ${\ov q}^1, q_1$ with masses $\omp\sim m_Q\mph/\la$ \eqref{(9.1)}, and similarly for hybrids made from $({\ov q}^{\,1} q_2)$ or $({\ov q}^{\,2} q_1)$.\\
2) A large number of hadrons made from weakly interacting non-relativistic quarks ${\ov q}^2, q_2$
with masses $\tmp\sim(\mo/\mph)^{N_c/(N_c-n_1)}\omp\ll\omp$, see \eqref{(9.1)}. All quarks are weakly confined, i.e. the tension of the confining string originating from ${\cal N}=1\,\, SU(\nd)$ SYM is $\sqrt\sigma\sim\lym^{(\rm br1)}\ll\tmp\ll\omp$.\\
3) A large number of gluonia from ${\cal N}=1\,\, SU(\nd)$ SYM, the scale of their masses is $\lym^{(\rm br1)}\sim(\langle M_1\rangle_{\rm br1}\langle M_2\rangle_{\rm br1}/\mph)^{1/3}$, see Appendix and \eqref{(9.1)}.\\
4) $n_2^2$ mions $M_2^2$ with masses $\mu^{\rm pole}(M_2^2)\ll\lym^{(\rm br1)}$, see \eqref{(9.5)}.\\
5) $n_1^2$ mions $M_1^1$ with masses $\mu^{\rm pole}(M_1^1)\ll\mu^{\rm pole}(M_2^2)$, see \eqref{(9.6)}.\\
6) $2n_1n_2$ mions $M_1^2, M_2^1$ are massless.

It is worth noting also that the scales $\mu_{\Phi,1}$ and $\mu_{\Phi,2}$,\,\, $\mu_{\Phi,2}\gg\mu_{\Phi,1}\gg\mo$, see \eqref{(8.4)},\eqref{(8.13)},\eqref{(8.20)} specific for the direct theory play no role in this dual theory at $1\leq n_1<{\rm Min}\,(N_F/2,\,3N_c-2N_F)$.\\

All formulas in special vacua with $\no=\nd,\, \nt=N_c$ can be obtained simply substituting $\no=\nd$ in all expressions in this section 9.1.

\subsection{\quad  br1 and special vacua with $3N_c-2N_F < n_1< N_F/2$}

The masses of dual quarks $\omp$ and $\tmp$ are the same (with a logarithmic accuracy), see \eqref{(9.1)}. The difference is that the dual coupling ${\ov a}(\mu)$ still decreased logarithmically with diminished scale at $\tmp<\mu<\omp$ at $1\leq n_1<{\rm Min}\,(N_F/2,\,3N_c-2N_F)$, while now $\bd^{\,\prime}=3\nd-n_2=n_1-
(3N_c-2N_F)>0$ and ${\ov a}(\mu)$ increases logarithmically with diminishing scale at $\mu<\omp$ until the dual theory enters the conformal regime at $\mu<{\tilde\Lambda}_Q$ (if nothing prevents). ${\tilde\Lambda}_Q$ is determined from the matching of dual couplings at $\mu=\omp$, see \eqref{(9.1)},\eqref{(8.18)},
\bq
({\tilde\Lambda}_Q )^{3\nd-n_2}=\la^{3\nd-N_F}(\omp)^{n_1}\quad\ra\quad {\tilde\Lambda}_Q=\la\Bigl (\frac{m_Q\mph}{\la^2}\Bigr )^{\frac{n_1}{n_1-(3N_c-2N_F)}\,>\, 0}=\Lambda^\prime_Q\ll\la\,.\label{(9.8)}
\eq

Therefore, we need $\tmp>\Lambda^\prime_Q$ in order not to enter the conformal regime, this requires, see
\eqref{(9.8)},\eqref{(9.1)},
\bq
\frac{\Lambda^\prime_Q}{\tmp}=\Bigl (\frac{\mph}{\mu_{\Phi,3}}\Bigr )^{\frac{n_1}
{(n_1-3N_c+2N_F)}\frac{2\nd}{(N_c-n_1)}}\ll 1\,\ra\, \mph\,\ll\,\mu_{\Phi,3} \label{(9.9)}
\eq
as in the direct theory, see \eqref{(8.20)}. The masses $\mu^{\rm pole}(M_2^2)$ and $\mu^{\rm pole}(M_1^1)\ll\mu^{\rm pole}(M_2^2)$ remain as in \eqref{(9.5)},\eqref{(9.6)} and besides, see \eqref{(9.5)},
\bq
\frac{\mu^{\rm pole}(M_2^2)}{\lym^{(\rm br1)}}\sim \Bigl (\frac{\mph}{\mu_{\Phi,3}}\Bigr )
^{\frac{4n_1}{3(N_c-n_1)}}\ll 1\quad{\rm at}\quad \mph\ll \mu_{\Phi,3}=\la\Bigl (\frac{\la}{m_Q}\Bigr )^{\frac{3N_c-2N_F+n_1}{2n_1}\,<\,1}<\frac{\la^2}{m_Q}\,.\label{(9.10)}
\eq

All formulas in special vacua with $\no=\nd,\, \nt=N_c$ can be obtained simply substituting $\no=\nd$ in all expressions in this section 9.2.

\subsection{\quad  br2 - vacua}

In dual br2 vacua with $N_F/2<\nt<N_c$, all formulae can also be obtained from those for dual br1-vacua by $\no\leftrightarrow\nt$, under corresponding restrictions.

\section{Conclusions}

\hspace*{6mm} This article continues our previous study in \cite{ch5,ch6,epj} of ${\cal N}=1$ SQCD-like theories with additional colorless but flavored fields. We considered this time the region $N_c<N_F<3N_c/2$. In this region, the UV free direct $SU(N_c)$ theory \eqref{(1.1)} with light quarks enters {\it smoothly} at $\mu<\la$ to the (very) strongly coupled regime with the coupling $a(\mu\ll\la)\gg 1$ (see Introduction, section 7 in \cite{ch1} and Conclusions in \cite{session} for additional arguments). The  mass spectra were calculated in its numerous different vacua. The calculations were performed within the dynamical scenario introduced in \cite{ch3}.  This scenario assumes that quarks in such ${\cal N}=1$ SQCD-like theories can be in two {\it standard} phases only\,: these are either the HQ (heavy quark) phase where they are confined, or the Higgs phase. The word {\it standard} implies here also that, in such theories without elementary colored adjoint scalars, no {\it additional}
\footnote{\,
i.e. in addition to the Nambu-Goldstone particles due to spontaneously broken global flavor symmetry
}
parametrically light solitons (e.g. magnetic monopoles or dyons) are formed at those lower scales $\mu\ll\la$ where quarks decouple as heavy or are higgsed.
\footnote{
Besides, it is worth noting that the appearance of additional parametrically light composite solitons will influence the 't Hooft triangles.
}

Similarly to our previous studies in \cite{ch5,ch6,epj} of these theories within the conformal window at $3N_c/2<N_F<3N_c$, it is shown here that, due to a strong powerlike RG evolution at scales $\mu<\la$ in the direct theory, the seemingly heavy and dynamically irrelevant fields $\Phi^j_i$ can become light and there appear then two additional generations of light $\Phi$-particles with $\mu_{2,3}^{\rm pole}(\Phi)\ll\la$.

In parallel, we calculated the mass spectra of IR free and logarithmically weakly coupled at $\mu<\la$ dual $\,SU(\nd=N_F-N_c)$ theory \eqref{(1.2)} proposed by Seiberg. Comparison have shown that mass spectra of the direct and dual theories are parametrically different, so that these two theories are not equivalent.

As it is seen from the article text, the use of the dynamical scenario from \cite{ch3} leads to the results for the mass spectra which look self-consistent. In
other words, no internal inconsistences appeared in all cases considered. It is worth to recall also that this dynamical scenario used in this article satisfies all those tests which were used as checks of the Seiberg proposal about the equivalence of the direct and dual theories. The parametric difference of mass spectra of direct and dual theories shows, in particular, that all these tests, although necessary, may well be insufficient.

{\appendix
\section{Condensates and multiplicities of vacua at $N_c<N_F<2N_c$}}

\hspace*{4mm} For the reader convenience and to make the present paper self-contained, we reproduce in short in this Appendix the values of quark and gluino condensates of the $SU(N_c)$ theory, $\langle{\ov Q}_j Q^i\rangle$ and $\langle S\rangle$, from the section 3 in \cite{ch4}.

As explained e.g. in section 3 of \cite{ch4} and/or in section 4 of \cite{epj}, the values of quark condensates in various vacua can be obtained from the {\it exact} effective superpotential depending on the quark bilinears $\Pi^i_j=({\ov Q}_j Q^i)$ only,
\footnote{\,
Recall that this is {\it not} a genuine low energy superpotential. \eqref{(A.1)} can be used {\it only} for finding the values of mean vacuum values
$\langle {\ov Q}_j Q^i\rangle$ and $\langle S\rangle$, see \cite{ch4}. The genuine low energy superpotentials in each vacuum are given in the text.
}
\bq
{\cal W}^{\,\rm eff}_{\rm tot}(\Pi)=m_Q{\rm Tr}\,({\ov Q}Q) -\frac{1}{2\mph}\Biggl [{\rm Tr}\,
({\ov Q}Q)^2- \frac{1}{N_c}({\rm Tr}\,{\ov Q}Q)^2  \Biggr ]+{\cal W}_{\rm non-pert}\,,\label{(A.1)}
\eq
\bbq
{\cal W}_{\rm non-pert}=-\nd S\,,\quad S=\Bigl (\frac{\det {\ov Q}Q}{\la^{\bo}}\Bigr )^{1/\nd}\,,\quad \bo=3N_c-N_F\,,\quad m_Q\ll\la\ll\mph\,.
\eeq

\subsection{The region $\la\ll\mph\ll\mo=\la(\la/m_Q)^{(2N_c-N_F)/N_c}$}

\subsubsection{Vacua with the unbroken flavor symmetry}

There are two groups of such vacua with parametrically different values of condensates, $\langle{\ov Q}_j Q^i\rangle_L=\delta^i_j\langle{\ov Q}Q\rangle_L$ and $\langle{\ov Q}_j Q^i\rangle_S=\delta^i_j\langle{\ov Q} Q\rangle_S$.

{\bf a}) There are $(2N_c-N_F)$ L-vacua (L=large) with
\bq
\qq_L\sim \la^2\Biggl (\frac{\la}{\mph}\Biggr )^{\frac{\nd}{2N_c-N_F}}\ll \la^2\,,\quad \langle\lym^{(L)}\rangle^3\equiv\langle S\rangle_{L}\sim\la^3\Biggl (\frac{\la}{\mph}\Biggr )^{\frac{N_F}{2N_c-N_F}}\,,\quad \nd=N_F-N_c\,.\label{(A.2)}
\eq
In these quantum L-vacua, the second term in the superpotential \eqref{(A.1)} gives numerically only a small correction.

{\bf b}) There are $(N_F-N_c)$ classical S-vacua (S=small) with
\bq
\qq_S\simeq -\,\frac{N_c}{\nd}\, m_Q\mph\,,\quad \langle\lym^{(S)}\rangle^3\equiv\langle S\rangle_{S}\sim\la^3\Biggl (\frac{m_Q\mph}{\la^2}\Biggr )^{\frac{N_F}{N_F-N_c}}\,.\label{(A.3)}
\eq
In these S-vacua, the first nonperturbative term in the superpotential \eqref{(A.1)} gives only small corrections with $Z_{N_F-N_c}$ phases, but just these corrections determine the multiplicity of these $\nd=(N_F-N_c)$ nearly degenerate vacua. On the whole, there are
\bq
N_{\rm unbrok}=(2N_c-N_F)+(N_F-N_c)=N_c \label{(A.4)}
\eq
vacua with the unbroken flavor symmetry at $N_c<N_F<2N_c$.

It follows from \eqref{(A.1)} that at $\mph\gg \mo$ the above $(2N_c-N_F)$ L - vacua and $(N_F-N_c)$ S - vacua degenerate into $N_c$ SQCD vacua
\bq
\langle{\ov Q}_j Q^i\rangle_{SQCD}\simeq\delta^i_j\frac{1}{m_Q}\Bigl (\lym^{(\rm SQCD)}\Bigr )^3=\delta^i_j\frac{1}{m_Q}\Bigl (\la^{\bo}m_Q^{N_F}\Bigr)^{1/N_c}\,,\quad \bo=3N_c-N_F\,.\label{(A.5)}
\eq

The value of $\mo$ is determined from the matching
\bbq
\Biggl [\langle\qq\rangle_L\sim \la^2\Biggl (\frac{\la}{\mo}\Biggr )^{\frac{\nd}{2N_c-N_F}}\Biggr ]\sim \Biggl [\langle\qq\rangle_S\sim m_Q\mo\Biggl ]\sim \Biggl [\langle\qq\rangle_{\rm SQCD}\sim \la^2\Bigl (\frac{m_Q}{\la}\Bigr )^{\frac{\nd}{N_c}}\Biggl ]\quad\ra
\eeq
\bq
\ra \mo\sim \la\Bigl (\frac{\la}{m_Q}\Bigr )^{\frac{2N_c-N_F}{N_c}}\gg \la\,.\label{(A.6)}
\eq

\subsubsection{Vacua with the spontaneously broken flavor symmetry, $U(N_F)\ra U(n_1)\times U(n_2)$}

In these, there are $n_1\leq [N_F/2]$ equal condensates $\langle{\ov Q}_1Q^1(\mu=\la)\rangle_{\rm br}\equiv\langle\Qo\rangle_{\rm br}$ and $n_2\geq n_1$ equal condensates $\langle{\ov Q}_2 Q^2(\mu=\la)\rangle_{\rm br}\equiv\langle\Qt\rangle_{\rm br}\neq\langle\Qo\rangle_{\rm br}$. The simplest way to find the values of quark condensates in these vacua is to use the Konishi anomalies \cite{Konishi}. These can be written as \cite{ch4}
\bbq
\langle\, \Qo+\Qt-\frac{1}{N_c}{\rm Tr}\,({\ov Q}Q)\, \rangle_{\rm br}=m_Q\mph\,,
\eeq
\bq
\langle S\rangle_{\rm br}=\Bigl(\frac{\det\qq_{\rm br}=\langle\Qo\rangle^{\rm {n}_1}_{\rm br}\langle\Qt\rangle^{\rm {n}_2}_{\rm br}}{\la^{\bo}}\Bigr )^{1/\nd}=\frac{\langle\Qo\rangle_{\rm br}\langle\Qt\rangle_{\rm br}}{\mph}\,,\label{(A.7)}
\eq
\bbq
\langle m^{\rm tot}_{Q,1}\rangle_{\rm br}\equiv\langle m_Q-\Phi_1\rangle_{\rm br}=\frac{\langle\Qt\rangle_{\rm br}}{\mph}\,,\quad \langle m^{\rm tot}_{Q,2}\rangle_{\rm br}\equiv\langle m_Q-\Phi_2\rangle_{\rm br}=\frac{\langle\Qo\rangle_{\rm br}}{\mph}\,.
\eeq
Besides, the multiplicity of vacua will be shown below at given values of $n_1$ and $n_2\geq n_1$.\\

{\bf a)} At $n_2\lessgtr N_c$, including $n_1=n_2=N_F/2$ for even $N_F$ but excluding $n_2=N_c$\,, there are
\footnote{\,
${\ov C}^{\,n_1}_{N_F}$ differ from the standard $C^{\,n_1}_{N_F}=(N_F!/{\rm n}_1!\,{\rm n}_2!)$ only by ${\ov C}^{\,n_1={\rm k}}_{N_F=2{\rm k}}=C^{\,n_1={\rm k}}_{N_F=2{\rm k}}/2$.
}
$(2N_c-N_F){\ov C}^{\,n_1}_{N_F}$ Lt-vacua (Lt=L-type) with the parametric behavior of condensates
\bq
(1-\frac{n_1}{N_c})\langle\Qo\rangle_{\rm Lt}\simeq -(1-\frac{n_2}{N_c})\langle\Qt\rangle_{\rm Lt}\sim \la^2\Biggl (\frac{\la}{\mph}\Biggr )^{\frac{\nd}{2N_c-N_F}}, \label{(A.8)}
\eq
i.e. as in the L-vacua above but $\langle\Qo\rangle_{\rm Lt}\neq\langle\Qt\rangle_{\rm Lt}$ here.

{\bf b)} At $n_2>N_c$ there are $(n_2-N_c)C^{{\rm n}_1}_{N_F}$ $\rm br2$-vacua (br2=breaking with the dominant $\langle\Qt\rangle$)
\bq
\langle\Qt\rangle_{\rm br2}\simeq\frac{N_c}{N_c-\nt} m_Q\mph\,,\quad \langle\Qo\rangle_{\rm br2}\sim \la^2
\Bigl (\frac{\mph}{\la}\Bigr )^{\frac{n_2}{n_2-N_c}}\Bigl (\frac{m_Q}{\la}\Bigr )^{\frac{N_c-n_1}{n_2-N_c}}\,,\label{(A.9)}
\eq
\bbq
\frac{\langle\Qo\rangle_{\rm br2}}{\langle\Qt\rangle_{\rm br2}}\sim \Bigl (\frac{\mph}{\mo}\Bigr )^{\frac{N_c}{n_2-N_c}}\ll 1\,,\quad \langle\lym^{(\rm br2)}\rangle^3\equiv\langle S\rangle_{\rm br2}\sim\la^3\Bigl (\frac{m_Q}{\la}\Bigr )^{\frac{n_2-n_1}{n_2-N_c}}\Bigl (\frac{\mph}{\la}\Bigr )^{\frac{n_2}{n_2-N_c}}\,.
\eeq

{\bf c)} At $n_1=\nd,\, n_2=N_c$ there are $(2N_c-N_F) C^{n_1=\nd}_{N_F}$ 'special' vacua with
\bq
\langle\Qo\rangle_{\rm spec}=\frac{N_c}{2N_c-N_F}(m_Q\mph)\,,\quad \langle\Qt\rangle_{\rm spec}=\la^2\Bigl (\frac{\la}{\mph}\Bigr )^{\frac{\nd}{2N_c-N_F}},\,\,\label{(A.10)}
\eq
\bbq
\frac{\langle\Qo\rangle_{\rm spec}}{\langle\Qt\rangle_{\rm spec}}\sim\Bigl (\frac{\mph}{\mo}\Bigr )^{\frac{N_c}{2N_c-N_F}}\ll 1\,,
\quad\langle S\rangle_{\rm spec}=\la^2 m_Q\Bigl (\frac{\la}{\mph}\Bigr )^{\frac{\nd}{2N_c-N_F}}\,.
\eeq

On the whole, there are (\,$\theta(z)$ is the step function\,)
\bq
N_{\rm brok}(n_1)=\Bigl [(2N_c-N_F)+\theta(n_2-N_c)(n_2-N_c)\Bigr ]{\ov C}^{\,n_1}_{N_F}= \label{(A.11)}
\eq
\bbq
=\Bigl [(N_c-\nd)+\theta(\nd-n_1)(\nd-n_1)\Bigr ]{\ov C}^{\,n_1}_{N_F}\,,
\eeq
vacua with the broken flavor symmetry $U(N_F)\ra U(n_1)\times U(n_2)$, this agrees with \cite{CKM}.

\subsection {The region $\mph\gg\mo$}.

{\bf a)} At all values of $n_2\lessgtr N_c$, including $n_1=n_2=N_F/2$ at even $N_F$ and the `special' vacua with $n_1=\nd,\, n_2=N_c$, there are $(N_c-n_1){\ov C}^{\,{\rm n}_1}_{N_F}$ $\rm br1$-vacua with
\bq
\langle\Qo\rangle_{\rm br1}\simeq\frac{N_c}{N_c-\no} m_Q\mph\,,\quad \langle\Qt\rangle_{\rm br1}\sim \la^2\Bigl (\frac{\la}{\mph}
\Bigr )^{\frac{n_1}{N_c-n_1}}\Bigl (\frac{m_Q}{\la}\Bigr )^{\frac{\nd-n_1}{N_c-n_1}}\,,\label{(A.12)}
\eq
\bbq
\frac{\langle\Qt\rangle_{\rm br1}}{\langle\Qo\rangle_{\rm br1}}\sim \Bigl (\frac{\mo}{\mph}\Bigr )^{\frac{N_c}{N_c-n_1}}\ll 1\,,\quad \langle\lym^{(\rm br1)}\rangle^3\equiv\langle S\rangle_{\rm br1}\sim\la^3\Bigl (\frac{m_Q}{\la}\Bigr )^{\frac{n_2-n_1}{N_c-n_1}}\Bigl (\frac{\la}{\mph}\Bigr )^{\frac{n_1}{N_c-n_1}}\,.
\eeq

{\bf b)} At $[N_F/2]\leq n_2<N_c$, there are also $(N_c-n_2){\ov C}^{\,n_2}_{N_F}=(N_c-n_2){\ov C}^{\,n_1}_{N_F}$\,\, $\rm br2$ - vacua with
\bq
\langle\Qt\rangle_{\rm br2}\simeq\frac{N_c}{N_c-\nt} m_Q\mph\,,\quad \langle\Qo\rangle_{\rm br2}\sim \la^2\Bigl (\frac{\la}{\mph}
\Bigr )^{\frac{n_2}{N_c-n_2}}\Bigl (\frac{\la}{m_Q}\Bigr )^{\frac{N_c-n_1}{N_c-n_2}}\,,\label{(A.13)}
\eq
\bbq
\frac{\langle\Qo\rangle_{\rm br2}}{\langle\Qt\rangle_{\rm br2}}\sim \Bigl (\frac{\mo}{\mph}\Bigr )^{\frac{N_c}{N_c-n_2}}\ll 1\,,\quad\langle\lym^{(\rm br2)}\rangle^3\equiv\langle S\rangle_{\rm br2}\sim\la^3\Bigl (\frac{\la}{m_Q}\Bigr )^{\frac{n_2-n_1}{N_c-n_2}}\Bigl (\frac{\la}{\mph}\Bigr )^{\frac{n_2}{N_c-n_2}}\,.
\eeq

On the whole, there are
\bq
N_{\rm brok}(n_1)=\Bigl [(N_c-n_1)+\theta (N_c-n_2)(N_c-n_2)\Bigr ]{\ov C}^{\,n_1}_{N_F}= \label{(A.14)}
\eq
\bbq
=\Bigl [(N_c-\nd)+\theta (\nd-n_1)(\nd-n_1)\Bigr ]{\ov C}^{\,n_1}_{N_F}
\eeq
vacua. As it should be, the number of vacua at $\mph\lessgtr \mo$ is the same.\\

As one can see from the above, all quark condensates become parametrically the same at $\mph\sim\mo=\la(\la/m_Q)^{(2N_c-N_F)/N_c}$. Clearly, this region $\mph\sim\mo$ is very special and most of the quark condensates change their parametric behavior and hierarchies at $\mph\lessgtr\mo$. For example, the br2-vacua with $n_2<N_c\,,\,\,\langle\Qt \rangle\sim m_Q\mph\gg\langle\Qo\rangle$ at $\mph\gg\mo$ evolve into the Lt (L-type) vacua with $\langle\Qt\rangle\sim\langle\Qo\rangle\sim \la^2 (\la/\mph)^{\nd/(2N_c-N_F)}$ at $\mph\ll\mo$, while the br2-vacua with $n_2>N_c\,,\,\,\langle\Qt\rangle\sim m_Q\mph\gg\langle\Qo\rangle$ at $\mph\ll\mo$ evolve into the br1-vacua with $\langle\Qo\rangle\sim m_Q\mph\gg\langle\Qt\rangle$ at $\mph\gg\mo$, etc. The exception is the special vacua with $n_1=\nd,\, n_2=N_c$\,. In these, the parametric behavior $\langle\Qo\rangle= m_Q\mph, \,\langle\Qt\rangle= \la^2(\la/\mph)^{\nd/(2N_c-N_F)}$ remains the same and only the hierarchy is reversed at $\mph\lessgtr\mo\, :\, \langle\Qo\rangle/\langle\Qt\rangle\sim (\mph/\mo)^{N_c/(2N_c-N_F)}$.\\

The total number of all vacua at $N_c<N_F<2N_c$ is
\bq
N_{\rm tot}=\Bigl ( N_{\rm unbrok}=N_c \Bigr )+\Bigl ( N_{\rm brok}^{\rm tot}=\sum_{n_1=1}^{[N_F/2]}N_{\rm brok}(n_1)\Bigr )=\sum_{k=0}^{N_c}(N_c-k)C^{\,k}_{N_F}\,,\label{(A.15)}
\eq
this agrees with \cite{CKM}\,.

The analog of \eqref{(A.1)} in the dual theory with $|\Lambda_q|=\la$ is obtained by the replacement ${\ov Q} Q(\mu=\la)\ra M(\mu=\la)$, $\langle M(\mu=\la)\rangle=\langle {\ov Q} Q(\mu=\la)\rangle$ in all vacua.  The multiplicities of different vacua are the same in the direct and dual theories.

\end{document}